\documentclass{LMCS}

\def\dOi{10(2:17)2014}
\lmcsheading%
{\dOi}
{1--24}
{}
{}
{Feb.~29, 2012}
{Jun.~30, 2014}
{}


\ACMCCS{[{\bf Software and its engineering}]: Software organization
  and properties---Software functional properties---Formal
  methods---Software verification}

\usepackage{amsmath, amssymb}

\usepackage{hyperref}

\theoremstyle{plain}

\theoremstyle{plain}
\newtheorem{theorem}[thm]{Theorem}

\theoremstyle{definition}

\renewcommand{\dotsm}{{\dots}}
\renewcommand{\dotsc}{{\ldots}}

\newcommand{\mcal}{\mathcal{M}}
\newcommand{\A}{\mathcal{A}}
\newcommand{\B}{\mathcal{B}}
\renewcommand{\H}{\mathcal{H}}
\newcommand{\M}{\mathsf{M}}
\renewcommand{\L}{\mathcal{L}}
\newcommand{\T}{\mathcal{T}}

\providecommand{\tuple}[1]{\langle#1\rangle}

\providecommand{\set}[1]{\{#1\}}

\begin{document}

\title{Synthesis from Probabilistic Components}

\author[Y.~Lustig]{Yoad Lustig\rsuper a}
\address{Department of Computer Science, Rice University,
  Houston, TX 77005, USA}
\email{yoad.lustig@gmail.com, \{nain,vardi\}@cs.rice.edu}

\author[S.~Nain]{Sumit Nain\rsuper b}	
\address{\vspace{-18 pt}}	%required
%\email{nain@cs.rice.edu}  %optional
%\thanks{thanks 1, optional.}	%optional

\author[M.~Y.~Vardi]{Moshe Y.~Vardi\rsuper c}	%optional
\address{\vspace{-18 pt}}	%optional
%\email{vardi@cs.rice.edu}  %optional
\thanks{{\lsuper c}Work supported in part by NSF grants CNS 1049862 and CCF-1139011,
by NSF Expeditions in Computing project ``ExCAPE: Expeditions in Computer
Augmented Program Engineering'', by BSF grant 9800096, and by gift from
Intel.}	%optional

%\author[]{Author 3}	%optional
%\address{address 3}	%optional
%\email{author3@email3}  %optional
%\thanks{thanks 3, optional.}	%optional

%% etc.

%% required for running head on odd and even pages, use suitable
%% abbreviations in case of long titles and many authors:

%% mandatory lists of keywords and classifications:
%\titlecomment{OPTIONAL comment concerning the title, if a variant
%or an extended abstract of the paper has appeared elsewehere}
%%%%%%%%%%%%%%%%%%%%%%%%%%%%%%%%%%%%%%%%%%%%%%%%%%%%%%%%%%%%%%%%%%%%%%%%%%%

%% the abstract has to PRECEED the command \maketitle:
%% be sure not to issue the \maketitle command twice!

\begin{abstract}
\noindent Synthesis is the automatic construction of a system from its
specification. In classical synthesis algorithms,
it is always assumed that the system is ``constructed from scratch''
rather than composed from reusable components. This, of course, rarely 
happens in real life, where almost every non-trivial commercial
software system relies heavily on using libraries of reusable
components. Furthermore, other contexts, such as web-service 
orchestration, can be modeled as synthesis of a system from a 
library of components.
Recently, Lustig and Vardi introduced \emph{dataflow} and
\emph{control-flow} synthesis from libraries of reusable
components. They proved that dataflow synthesis is undecidable,
while control-flow synthesis is decidable.
In this work, we consider the problem of control-flow synthesis from
libraries of \emph{probabilistic components}. We show that this more 
general problem is also decidable.
\end{abstract}

\maketitle

\section{Introduction}
Hardware and software systems are rarely built from scratch. Almost
every non-trivial system is based on existing components. A typical
component might be used in the design of multiple systems.
Examples of such components include function libraries,
web APIs, and ASICs. Consider the mapping application in a typical
smartphone. Such an application might call the location service
provided by the phone's operating system to get the user's co-ordinates, 
then call a web API to obtain the correct map image tiles, and finally 
call a graphics library to display the user's location on the screen. 
None of these components are exclusive to the mapping application and
all of them are commonly used by other applications. 

The construction of systems from reusable components is an area of active 
research.  Examples of important work on the subject can be found in 
Sifakis' work on component-based construction \cite{Sif05}, and  de Alfaro 
and Henzinger's work on ``interface-based design'' \cite{dAH05}.  
Furthermore, other situations, such as web-service orchestration 
\cite{BCGLM03}, can be viewed as the construction of systems 
from libraries of reusable components.

Synthesis is the automated construction of a system from its specification. 
In contrast to model checking, which involves verifying that a system 
satisfies the given specification, synthesis aims to automatically 
construct the required system from its formal specification. 
The modern approach to temporal synthesis was initiated by Pnueli and 
Rosner who introduced linear temporal logic (LTL) synthesis \cite{PR89a}.  
In LTL synthesis, the specification is given in LTL and the system 
constructed is a finite-state transducer modeling a reactive system.  In 
this setting it is always assumed that the system is 
``constructed from scratch'' rather than ``composed'' from 
existing components.  Recently, Lustig and Vardi \cite{LV09} 
introduced the study of synthesis from reusable components. The use of 
components  abstracts much of the detailed behavior of a 
sub-system, and allows one to write specifications that mention only
the aspects of sub-systems relevant for the synthesis of the system at large.
 
A major concern in the study of synthesis from reusable components is the 
choice of a mathematical model for the components and their composition. 
The exact nature of the reusable components in a software library may 
differ.  One finds in the literature 
many different types of components; for example, function libraries 
(for procedural programming languages) or object libraries (for 
object-oriented programming languages).  Indeed, there is no single 
``right'' model encompassing all possible facets of the problem. The 
problem of synthesis from reusable components is a general problem to 
which there are as many facets as there are models for components and 
types of composition \cite{Sif05}. 
 
As a basic model for a component, following \cite{LV09}, we abstract away 
the precise details of the component and model a component as a 
{\em transducer}, i.e., a finite-state machine with 
outputs. Transducers constitute a canonical model for reactive 
components, abstracting away internal architecture and focusing on 
modeling input/output behavior.  In \cite{LV09}, two models of 
composition were studied. In \emph{dataflow} composition, 
the output of one component is fed as input to another component. 
The synthesis problem for dataflow composition was shown to be 
undecidable.  In {\em control-flow} composition 
control is held by a single component at every point in time. The 
synthesis problem can then be viewed as constructing a supervisory 
transducer that switches control between the component 
transducers. Control-flow composition is motivated by 
software (and web services) in which a single function is in control 
at every point during the execution. LTL synthesis in this setting was 
shown in \cite{LV09} to be 2EXPTIME-complete, just like classical LTL 
synthesis \cite{PR89a}. 
 
In this paper, we extend the control-flow synthesis model of 
\cite{LV09} to probabilistic components, which are transducers with a 
probabilistic transition function. This is a well known approach 
to modeling systems where there is probabilistic uncertainty about 
the results of input actions. 
Intuitively, we aim at constructing a reliable system from 
unreliable components.  There is a rich literature about 
verification and analysis of such systems, 
cf.~\cite{Var85b,CY90,CY95,Var99}, 
as well about synthesis in the face of probabilistic uncertainty 
\cite{BGLBC04}.  The introduction of probability requires us to use a 
probabilistic notion of correctness; here we choose the 
\emph{qualitative} criterion that the specification be satisfied with 
probability~$1$, leaving the study of \emph{quantitative criteria} to 
future work. 
%myv1: This sentence is confusing. In the full version, we'll need
%to add a longer discussion of this.
%In general, deterministic synthesis can also be viewed as a 2 player game.
%In this setting, our work can be considered as the 2-1/2 player 
%generalization of the component based synthesis of Lustig and Vardi 
%\cite{LV09}.

Here, our focus is on proving decidability, rather than on establishing 
precise complexity bounds, leaving the study of precise bounds to future 
work.  Consequently, we abstract away from the details 
of the specification formalism and assume that the specification is 
given in terms of deterministic parity word automata (DPW). This allows us 
to consider all $\omega$-regular properties.  We define and study the 
\emph{DPW probabilistic realizability} and synthesis problems, where the 
input is a library $\L$ of probabilistic components and a 
DPW $\mathcal{A}$, and the question is whether one can construct a 
\emph{finite} system $S$ from the components in $\L$,  such that, 
regardless of the external environment, the traces generated by the 
system~$S$ are accepted by~$\mathcal{A}$ with probability~1. Each
component in the library can be used an arbitrary number of times in
the construction and there is no apriori bound on the size of the
system obtained. The technical challenge here is dealing with the
finiteness of the system under construction.
In \cite{LV09}, as well as in \cite{PR89a}, one need not 
deal with finiteness from the start. In fact, one can test realizability 
without being concerned with finiteness of the constructed system, 
as finiteness is a \emph{consequence} of the construction. This is not the 
case here, where we need to deal with finiteness from the start. 
Nevertheless, we are able to show that the problem is in 2EXPTIME. 
 
Before tackling the full problem, we first consider a restricted 
version of the problem, where the specification is given in the form 
of a parity index on the states of the components, and the composed 
system must satisfy the parity condition. We call this the 
\emph{embedded parity realizability} problem. We solve this 
problem and then show how solving the embedded parity realizability 
problem directly allows us to solve the more general DPW probabilistic 
realizability problem as well. The key idea here is that by taking the 
product of the specification DPW with each of the components, we can 
obtain larger components each of whose states has a parity associated 
with it. The challenge in completing the reduction is the need to 
generate a static composition, which does not depend on the 
history of the computation. Here we use ideas about synthesis with 
incomplete information from \cite{KV97c}. 

%\noindent\colorbox{lightgray}{\parbox{\columnwidth}{
%The paper is self-contained, except for certain technical proofs that
%have been omitted to save space; a longer version is posted on the
%authors' home pages.

\section{Background}

\subsection{Preliminaries}

\subsubsection{Labeled Trees}
%\textbf{\emph{Labeled Trees}}:
Given a set $D$ of directions, a {\em $D$-tree} is a set
$T \subseteq D^\ast$ such that  (a) there is an element $x_0 \in T$, called the \emph{root} of T, such that, for all $x\in T$ there exists $y\in D^\ast$ with $x = x_0\cdot y$, and (b) 
if $x \cdot c$ is a non-root element of $T$, where
$x \in D^\ast$ and $c \in D$, then $x$ is also an element of $T$. The elements of $T$ are called its \emph{nodes}.
For every node $x \in T$, the set of {\em successors} of $x$ is given by $\{x \cdot c \in T : c\in D\}$. A node with no successors is called a \emph{leaf}.
A {\em path} $\pi$ of a tree $T$ is a set $\pi \subseteq T$ such for every pair of nodes $x, y$ in $\pi$, there exists $z\in D^\ast$ such that $x = y\cdot z$ or $y = x\cdot z$. A path is infinite if it has no leaf nodes, otherwise it is finite. A \emph{subtree} of $T$ is a tree $T'\subseteq T$. For a node $x\in T$, the \emph{subtree rooted at $x$}
% , denoted $T(x)$, 
is the tree $\{x\cdot y \in T : y\in D^\ast\}$.
The {\em full} $D$-tree is $D^*$. The \emph{full subtree} at $x$ is the tree
whose set of nodes is $x\cdot D^\ast$. 

Given an alphabet $\Sigma$, a {\em
  $\Sigma$-labeled $D$-tree} is a pair 
$\tuple{T,\tau}$, where $T$ is a tree and $\tau:T \rightarrow \Sigma$ maps each
node of $T$ to a letter in $\Sigma$. A \emph{subtree} of
$\tuple{T, \tau}$, is a $\Sigma$-labeled $D$-tree $\tuple{T', \tau'}$, 
where $T'$ is a subtree of $T$ and $\tau'(x) = \tau(x)$, for all $x\in T'$.

%Given a set $D$ of directions, a {\em $D$-tree} is a set
%$T \subseteq D^{*}$ such that if $x \cdot c \in T$, where
%$x \in D^{*}$ and $c \in D$, then also $x \in T$.
%For every $x \in T$, the words $x \cdot c$, for $c \in D$, are the
%{\em successors} of $x$.
%A {\em path} $\pi$ of a tree $T$ is a set $\pi \subseteq T$ such that
%$\varepsilon \in \pi$ and for every $x \in \pi$, either $x$ is a leaf or
%there exists a unique $c \in D$ such that $x \cdot c \in \pi$.
%The {\em full} $D$-tree is $D^*$. Given an alphabet $\Sigma$, a {\em
%  $\Sigma$-labeled $D$-tree} is a pair 
%$\tuple{T,\tau}$, where $T$ is a tree and $\tau:T \rightarrow \Sigma$ maps each
%node of $T$ to a letter in $\Sigma$. A \emph{subtree} of
%$\tuple{D^\ast, \tau}$, is a $\Sigma$-labeled $D$-tree $\tuple{T,
%  \tau'}$, where $\tau'(x) = \tau(x)$, for all $x\in T$. For a
%  node $x\in D^\ast$, the \emph{full subtree} at $x$ is the subtree
%  whose set of nodes is $x\cdot D^\ast$.

\subsubsection{Tree Automata}
%\textbf{\emph{Tree Automata}}:
For a set $X$, let $\mathcal{B^+}(X)$ be the set of positive Boolean formulas over $X$ (i.e., Boolean
formulas built from elements in $X$ using $\wedge$ and $\vee$), including the formulas
\textbf{True} (an empty conjunction) and \textbf{False} (an empty disjunction). For a set $Y\subseteq X$ and
a formula $\theta\in \mathcal{B}^{+}(X)$, we say that $Y$ \emph{satisfies} $\theta$ iff assigning \textbf{True} to elements in $Y$
and assigning \textbf{False} to elements in $X-Y$ makes $\theta$ true. An \emph{alternating tree automaton}
is tuple $\A = \tuple{\Sigma, D, Q, q_0, \delta, \beta}$ , where $\Sigma$ is the input alphabet, $D$ is a set of directions, $Q$
is a finite set of states, $q_0\in Q$ is an initial state, $\delta:  Q\times \Sigma \rightarrow \mathcal{B}^{+}(D\times Q)$ 
 is a transition function, and $\beta$ specifies the acceptance condition that defines a subset of $Q^\omega$. 
Each element of $\mathcal{B}^{+}(D\times Q)$ is called an \emph{atom}.
%For a state q 2 Q, we denote by Aq the automaton h; D; Q; q; ; i in which q is the initial state.
The alternating automaton $\A$ runs on $\Sigma$-labeled full $D$-trees. A run of $\A$ over a $\Sigma$-labeled $D$-tree $\tuple{T,\tau}$
is a $(T\times Q)$-labeled $\mathbb{N}$-tree $\tuple{T_r, r}$. Each node of $T_r$ corresponds
to a node of $T$. A node in $T_r$, labeled by $(x,q)$, describes a copy of the automaton that
reads the node $x$ of $T$ and visits the state $q$. Note that multiple nodes of $T_r$ can correspond
to the same node of $T$. The labels of a node and its successors have to satisfy the
transition function. Formally, $\tuple{T_r, r}$ satisfies the following conditions:
\begin{enumerate}
\item  $\epsilon\in T_r$ and $r(\epsilon) = (\epsilon, q_0)$.
\item Let $y\in T_r$ with $r(y) = (x,q)$ and $\delta(q,\tau(x)) = \theta$. Then there exists a set $S = \{(c_0,q_0),(c_1,q_1),\dotsc,(c_n,q_n)\} \subseteq D\times Q$ such that $S$ satisfies $\theta$,
and for all $0\leq i\leq n$, we have $y\cdot i\in T_r$ and $r(y\cdot i) = (x\cdot c_i, q_i)$. $S$ is allowed to be empty.
\end{enumerate}
An infinite path $\pi$ of a run $\tuple{T_r, r}$ is labeled by a word in $Q^\omega$. Let $inf(\pi)$ be the set of states in $Q$ that occur infinitely often in $r(\pi)$. The \emph{B\"{u}chi} acceptance condition is given as $\beta\subseteq Q$, and $\pi$ satisfies $\beta$ if $inf(\pi)\cap \beta \neq \emptyset$. The \emph{parity} acceptance condition is given as a function $\beta:Q\rightarrow \{1,\dotsc, k\}$, and $\pi$ satisfies $\beta$ if $\min(\{\beta(q): q\in inf(\pi)\})$ is even. 
A run $\tuple{T_r, r}$ is accepting if all its infinite paths satisfy the acceptance condition. 
%Given a run hTr; ri and an in?nite path   Tr, let inf()  Q be such that q 2 inf()
%if and only if there are in?nitely many y 2  for which r(y) 2 T  fqg. We consider
%Buchi � acceptance in which a path  is accepting iff inf() \  =6 ;, and co-Buchi �acceptance in which a path  is accepting iff inf() \  = ;. 
An automaton accepts a tree iff there exists a run that accepts it. We denote by $\mathcal{L}(\A)$ the set of all $\Sigma$-labeled $D$-trees accepted by $\A$.

The transition function $\delta$ of an alternating tree automaton is \emph{nondeterministic} if every formula produced by $\delta$ can be written in disjunctive normal form such that if two atoms $(c_1,q_1)$ and $(c_2,q_2)$ occur in the same conjunction then $c_1$ and $c_2$ must be different.
A \emph{nondeterministic tree automaton} $\A$ is an alternating tree automaton with a nondeterministic transition function. 
In this case the transition function returns a set of $|D|$-ary tuples of states and can be represented as a function $\delta: Q\times \Sigma \rightarrow 2^{Q^{|D|}}$.

%We denote each of the different types of automata by three-letter acronyms in fD; N; Ug fB; Cg  fW; Tg, 
%We consider two different accepting conditions: B\"{u}chi and parity. 
%where the ?rst letter describes the branching mode of the automaton (deterministic, nondeterministic, or universal), the second letter describes the acceptance condition (Buchi or co-B � uchi), and the third letter describes the object over �
%which the automaton runs (words or trees). For example, NBT are nondeterministic tree
%automata and UCW are universal co-Buchi word automata.

\subsubsection{Transducers}
%\textbf{\emph{Transducers}}:
A \emph{deterministic transducer} is a tuple $B=  \tuple{ \Sigma_I,
  \Sigma_O, Q, q_0, \delta, L}$, where: 
$\Sigma_I$ is a finite input alphabet,
$\Sigma_O$ is a finite output alphabet,
$Q$ is a finite set of states,
$q_0\in Q$ is an initial state,
$L:Q\to\Sigma_O$ is an output function labeling states with output
letters, and $\delta:Q\times\Sigma_I \to Q$ is a transition
function. We define $\delta^\ast: \Sigma_I^\ast \rightarrow Q$ as
follows: 
$\delta^\ast(\epsilon) = q_0$ and for $x\in \Sigma_I^\ast$ and $a\in \Sigma_I$,
$\delta^\ast(x\cdot a) = \delta(\delta^\ast(x), a)$. 
We denote by $tree(B)$,  the $\Sigma_O$-labeled $\Sigma_I$-tree
$\tuple{\Sigma_I^\ast, \tau}$, where for all $x\in \Sigma_I^\ast$, 
we have $\tau(x) = L(\delta^\ast(x))$. We say $tree(B)$ is
the \emph{unwinding} of $B$. A $\Sigma$-labeled $D$-tree $T$ is called
\emph{regular}, if there exists a deterministic transducer $C$ such that $T = tree(C)$.

A probability distribution on a finite set $X$ is a function $f:
X\rightarrow [0,1]$ such that $\sum_{x\in X} f(x) = 1$. We use
$Dist(X)$ to denote the set of all probability distributions on set
$X$.  A {\em probabilistic transducer},
is a tuple
$\T = \tuple{ \Sigma_I, \Sigma_O, Q, q_0, \delta, F, L}$, where:
$\Sigma_I$ is a finite input alphabet,
$\Sigma_O$ is a finite output alphabet,
$Q$ is a finite set of states,
$q_0\in Q$ is an initial state,
$\delta:(Q-F)\times\Sigma_I \to Dist(Q)$ is a probabilistic transition function,
$F\subseteq Q$ is a set of exit states, and
$L:Q\to\Sigma_O$ is an output function labeling states with output
letters. Note that there are no transitions out of an exit state. If $F$
is empty, we say $\T$ is a probabilistic transducer without exits. Note that deterministic transducers are a special case of probabilistic transducers.

Given a probabilistic transducer $M = (\Sigma_I, \Sigma_o, Q, q_0, \delta, F, L)$, a
\emph{strategy} for $M$ is a function $f: Q^\ast\rightarrow
Dist(\Sigma_I)$ that probabilistically 
chooses an input for each sequence of states. A strategy is memoryless if the
choice depends only on the last state in the sequence. A memoryless
strategy can be written as a function $g:Q\rightarrow
Dist(\Sigma_I)$. A strategy is \emph{pure} if the choice is
deterministic. A pure strategy is a function $h: Q^* \rightarrow
\Sigma_I$, and a memoryless and pure strategy is a function $h: Q\rightarrow
\Sigma_I$. 

A strategy $f$ along with a probabilistic transducer $M$,
with set of states $Q$, induces a probability distribution on
$Q^\omega$, denoted $\mu_f$. By standard measure theoretic arguments, it suffices to define $\mu_f$ for the
cylinders of $Q^\omega$, which are sets of the form $\beta \cdot Q^\omega$,
where $\beta \in Q^\ast$. First we extend $\delta$ to exit states as
follows: for $a\in \Sigma_I$, $q\in F$, $q' \in Q$,
$\delta(q,a)(q) = 1$ and $\delta(q,a)(q') = 0$ when $q' \neq q$. 
Then we define $\mu_f(q_0\cdot Q^\omega) = 1$, and for  $\beta \in
Q^\ast$, $q, q'\in Q$, $\mu_f(\beta qq'\cdot Q^\omega) = \mu_f(\beta
q)(\sum_{a\in \Sigma_I} f(\beta q)(a)\times \delta(q,a)(q'))$.
These conditions say that there
is a unique start state, and the probability of visiting a state $q'$,
after visiting $\beta q$, is the same as the probability of the strategy picking a particular
letter multiplied by the probability that  the transducer transitions
from $q$ to $q'$ on that input letter, summed over all input
letters. 
%The output mapping $L: Q\rightarrow \Sigma_O$ can be extended to
%to a probability distribution $\mu_f^L$ on $\Sigma_O^\omega$ 

\subsubsection{Graph Induced by a Strategy}
Given a directed graph $G = (V, E)$,  a \emph{strongly connected component} of $G$ is a subset $U$ of $V$, such that for all $u,v \in  U$, $u$ is reachable from $v$.
%A maximal strongly connected component (MSCC) is an SCC $P\subseteq
%Q$ such that if $P'\subseteq Q$ is another SCC, then $P-P'\neq
%\emptyset$. 
We can define a natural partial order on the set of maximal strongly connected components of $G$ as follows: $U_1\leq U_2$ if
there exists $u_1\in U_1$ and $u_2\in U_2$ such that $u_1$ is reachable
from $u_2$. Then $U\subseteq V$ is an \emph{ergodic set} of $G$ if it is a minimal
element of the partial order. 

Let $M$ be a probabilistic transducer, $Q$ be its set of states, and $f$ be a
memoryless strategy for $M$. We define the graph induced by $f$ on $Q$,
denoted by $G_{M,f}$, as the directed graph $(Q, E)$, where $(q_1,q_2)\in
E$ if $\sum_{a\in \Sigma_I}f(q_1)(a)\, \delta(q_1,a)(q_2) > 0$. That
is, there is an edge from $q_1$ to $q_2$ if the transducer can
transition from the state $q_1$ to the state $q_2$ on an input letter that the
strategy chooses with positive probability.
Given $q_1,q_2\in Q$, we say that $q_2$ is
reachable from $q_1$ if there is a path from $q_1$ to $q_2$ in $G_{M,f}$. 
We say a state is ergodic if it belongs
to some ergodic set of $G_{M,f}$. An ergodic set is reachable if there is a path
from the start state to some state in the ergodic set.
A state $q$ of $M$ is \emph{reachable under $f$}, if there is a
path in $G_{M,f}$ from $q_0$ to $q$. 

\subsubsection{Library of Components}
A {\em library} is a set of probabilistic transducers 
that share the same input and output alphabets. Each transducer in the
library is called a \emph{component}. Given a finite set of directions
$D$, we say a library $\L$ has width $D$, if each component in the
library has exactly $|D|$ exit states. Since we can always add dummy
unreachable exit states to any component, we assume, w.l.o.g., that
all libraries have an associated width, usually denoted $D$. In the
context of a particular component, we often refer to elements
of $D$ as exits, and subsets of $D$ as sets of exits. 
Given a component $M$ from library $\L$, and
a strategy $f$ for $M$, we say that the exit $i\in D$ is
\emph{selected} by $f$, if the $i$th exit state of $M$ is reachable
under $f$.

An \emph{index function} for a transducer is a function that assigns
a natural number, called a priority index, to each state of the transducer.
An index function for a library is a function that assigns a priority 
to every state of every component in the library. Given an index
function $\alpha$ for a library $\L$, we define $\max(\alpha)$ to be
the highest priority assigned by $\alpha$. We can assume, w.l.o.g., that
$\max(\alpha)$ is not larger than  twice the maximal number of
states in the components of the library. 
Given a transducer $M$, index function $\alpha$, and a strategy $f$
for $M$, we say $f$ \emph{visits} priority $p$ if there
exists a state $q$ of $M$ such that $\alpha(q) = p$ and $q$ is
reachable under $f$.

\subsection{Reactive Synthesis}
Reactive synthesis involves the automated construction of reactive programs from specifications.
Given sets $I$ and $O$ of input and output signals, respectively, we can view a program as a function $P: (2^I)^\ast \rightarrow 2^O$ that maps a finite sequence of sets of input signals into a set of output signals. 
A reactive system can be viewed as a non-terminating program that interacts with an adversarial environment.
The environment generates an infinite sequence of input signals, which are modeled as infinite words over the alphabet $2^I$. The execution of the program for a particular input word results in an infinite computation, which is represented as an infinite word over $2^{(I \cup O)}$.

Given an LTL formula $\psi$ over $I \cup O$, realizability of $\psi$ is the problem of determining whether there exists a program $P$ all of whose computations satisfy the specification $\psi$. The correct synthesis of $\psi$ then amounts to constructing such $P$ \cite{PR89a}.

The complete behavior of the system can be described by the set of all possible executions (i.e. the \emph{traces} of the system), which is represented as a  $2^O$-labeled $2^I$-tree, called an \emph{execution tree}. The automata-theoretic approach involves constructing a tree automaton that accepts all computation trees all of whose paths satisfy $\psi$. The solution to the LTL synthesis problem then consists of a reduction to the nonemptiness problem of tree automata \cite{PR89a} (an earlier and more complicated solution can be found in \cite{BL69}). The LTL synthesis problem  is closely related to Church's problem \cite{Chu63, Rab70}.

The automata-theoretic approach to synthesis has been quite fruitful since the original work of Pnueli and Rosner \cite{PR89a}. Automata-theoretic methods have been applied successfully to the synthesis of branching specifications \cite{KMTV00} and to synthesis in the presence of incomplete or hidden information \cite{KV97c}. The work reported in this paper extends the reactive-synthesis framework to synthesis from probabilistic components.

%\subsubsection{Synthesis from Deterministic Components}

%Though the program P is deterministic, it induces a computation tree. The branches of the tree correspond to external nondeterminism, caused by different possible inputs. Thus, the tree has a fixed branching degree |2I |, and it embodies all the possible inputs (and hence also computations) of P . 

%
%The linear paradigm for realizability and synthesis is closely related to Church�s solvability problem (Chu63). There, we are given a regular relation R ? (2I)? ? (2O)? and we seek a function f : (2I)? ? (2O)?, generated by a strategy, such that for all x ? (2I)?, we have R(x,f(x)). We can view the relation R as a linear specification for the program: it defines all the permitted pairs of input and output sequences. A function f as above then maps every possible input sequence into a permitted output sequence, and can be therefore viewed as a correct program. The solutions to Church�s problem and the LTL synthesis problem are similar (Rab70; PR89), and consist of a reduction to the nonemptiness problem of tree automata (an earlier and more complicated solution can be found in (BL69)).
%Though the program P is deterministic, it induces a computation tree. The branches of the tree correspond to external nondeterminism, caused by different possible inputs. Thus, the tree has a fixed branching degree |2I |, and it embodies all the possible inputs (and hence also computations) of P . 

\section{Control-flow Composition from Libraries}
We first informally describe our notion of control-flow
composition of components from a library.
The components in the
composition take turns interacting with the 
environment, and at each point in time, exactly one component is
active. When the active component reaches an exit state, control is
transferred to some other component. Thus, to define a control flow
composition, it suffices to name the components used and describe how
control should be transferred between them. We use a deterministic
transducer to define the transfer of control.
Each library component can be used multiple times in a composition, and we
treat these occurrences as distinct \emph{component instances}. 
We emphasize that the composition can contain
potentially arbitrarily many repetitions of each component inside
it.
 Thus, the size of the composition, a priori, is not bounded.
Note that our notion of composition is \emph{static}, where the components
called are determined before run time, rather than \emph{dynamic}, where 
the components called are determined during run time. 

Let $\L$ be a library with width $D$. A \emph{composer} over $\L$ is a
deterministic tranducer 
$C = (D, \L, \mathcal{M}, \M_0, \Delta, \lambda)$. Here $\mathcal{M}$
is an arbitrary finite set of states. There is no bound on the size of
$\mathcal{M}$. Each $\M_i\in \mathcal{M}$ is the name of an instance
of a  component from $\L$ and $\lambda(\M_i) \in \L$ is the type of 
$\M_i$. We use the following notational convention for component
instances and names: the upright letter $\M$ always denotes
component names (i.e. states of a composer) and the italicized letter
$M$ always denotes the corresponding component instances (i.e. elements of
$\L$). Further, for notational convenience we often write $M_i$
directly instead of $\lambda(\M_i)$. Note that while each $\M_i$ is
distinct, the corresponding components $M_i$ need not 
be distinct. 
Each composer defines a unique composition over components from
$\L$. The current state of the composer corresponds to the component
that is in control. The transition function $\Delta$ describes how to transfer
control between components: $\Delta(\M,i) = \M'$ denotes that when the
composition is in the $i$th  final state of component $M$ it moves to
the start state of component $M'$. 
A composer can be viewed as an implicit
representation of a composition. We give an explicit definition of 
composition below.
\begin{defi}[Control-flow Composition]\label{def:composition}
Let $C = (D,\L,\mathcal{M}, \M_0,\Delta,
\lambda)$ be a composer over library $\L$ with width $D$, where
$\mathcal{M} = \{\M_0,\dotsc,\M_n\}$, $\lambda(\M_i) = (\Sigma_I, \Sigma_O,
Q_i, q_0^i, \delta_i, F_i, L_i)$ and $F_i = \{q_x^i : x\in D\}$. The
composition defined by $C$, denoted $\T_C$, is a probabilistic
transducer 
 $\tuple{\Sigma_I,\Sigma_O, Q, q_0, \delta, \emptyset, L }$, where $Q
 =\bigcup_{i=0}^n (Q_i\times\{i\})$, $q_0 = \tuple{q^0_0,0}$, 
$L(\tuple{q,i}) = L_i(q)$, and the transition function $\delta$ is
defined as follows: For $\sigma\in \Sigma_I$, $\tuple{q,i}\in Q$ and  $\tuple{q',j} \in Q$,

\vspace{2pt}
\begin{enumerate}
\item If $q\in Q_i\setminus F_i$, then
\[
\qquad \delta(\tuple{q,i},\sigma)(\tuple{q',j}) =
\begin{cases}
\delta_i(q,\sigma)(q') &  \text{if $i = j$} \\
0 & \text{otherwise}
\end{cases}
\]
\item If $q = q_x^i \in F_i$, where  $\Delta(\M_i,x) = \M_k$, then 
\[
 \qquad \delta(\tuple{q,i},\sigma)(\tuple{q', j}) = 
\begin{cases}
1 & \text{if $j$ = $k$ and $q' = q_0^k$} \\
0 & \text{otherwise}
\end{cases}
\]
\end{enumerate}
\end{defi}

Note that the composition is a probabilistic transducer without exits.
When the composition is in a state $\tuple{q,i}$ corresponding to a non-exit
state $q$ of component $M_i$, it behaves like $M_i$. When the
composition is in a state $\tuple{q_f, i}$ corresponding to an exit state
$q_f$ of component $M_i$, the control is transferred to the start state
of another component as determined by the transition function of the
composer. Thus, at each point in time, only one component is active
and interacting with the environment.

\section{Synthesis for Embedded Parity}\label{sec:embedded-parity}

In this section we consider a simplified version of the general synthesis
problem, where each state of a component in the library has a priority
associated with it and the specification to be satisfied is that the
highest priority visited i.o. must be even with probability $1$.

Let $M$ be a probabilistic tranducer and $\alpha$ be an index
function. A strategy $f$ for $M$ is \emph{winning} for the
environment if with positive probability the highest priority visited
infinitely often (i.o.) is odd.
We say that $M$
\emph{satisfies} $\alpha$ if there exists no winning strategy for the
environment. Given a composer $C$ over library $\L$, we say that $C$
\emph{satisfies} $\alpha$ if $\T_C$ satisfies $\alpha$.

Given a library $\L$ with width $D$, an \emph{exit control relation} is a
set $R\subseteq D \times \L$. 
We say that a composer $C = (D, \L, \mathcal{M}, \M_0, \Delta,
\lambda)$ over $\L$ is \emph{compatible}
with $R$, if the following holds: for all $\M, \M'\in \mathcal{M}$ and
$i\in D$, if $\Delta(\M, i) = \M'$ then $(i,M') \in R$. Thus, each
element of $R$ can be viewed as a constraint on how the composer is
allowed to connect components.

\begin{defi}
The \emph{embedded parity realizability problem} is:
Given a library
$\L$ with width $D$, an exit control relation $R$ for $\L$,
 and an index function $\alpha$ for $\L$, decide whether there exists a
composer $C$ over $\L$, such that $C$ satisfies 
$\alpha$ and $C$ is compatible with $R$. If such a composer exists, we
say that $\L$ \emph{realizes} $\alpha$ under $R$.
The \emph{embedded parity synthesis problem} is to find such a
composer $C$ if it exists.
\end{defi}

The following theorem allows us to restrict attention to memoryless
strategies. It states that if a winning strategy exists, then a
memoryless winning strategy must also exist. Here we give a direct
combinatorial proof, but we note that the result can also be obtained
by adapting the methods in \cite{CJH03}, where a similar result was proved
for $2$--$1/2$ player stochastic parity games by Chatterjee et al.  
\begin{theorem}\label{theorem:memoryless}
Given a probabilistic transducer $M$, and index function $\alpha$, if there exists a winning strategy for the environment then there exists a pure and memoryless winning strategy.
\end{theorem}
\proof 
We break up the proof of this theorem in two parts in Lemma \ref{lemma:memoryless-strategy} and Lemma \ref{lemma:pure-strategy}. In the first part we show that given a winning strategy $f$ we can find a memoryless winning strategy $f'$ from $f$. In the second part we show that given a memoryless winning strategy $f'$, we can obtain a  pure and memoryless strategy $f''$ from $f'$. Together the two lemmas suffice to complete the proof.
\qed

% We will need the following lemma from probability theory:
% \begin{lem}
% Let $\mu$ be a probability measure on the set of infinite words over
% $Q$ and let $P\subseteq Q$. Then $\mu(\beta\cdot P^\omega) = 0$ or
% $\mu(\beta\cdot P^\omega)  = \mu(\beta\cdot Q^\omega)$.
% \end{lem}

\begin{lem} \label{lemma:memoryless-strategy}
Let $M$ be a transducer and $f$ be a winning strategy for the
environment. Then there exists a memoryless strategy $g$ such that
$g$ is winning.
\end{lem}

\proof 
Let $f$ be a strategy that is winning for the
environment. Let $Q$ be the set of states of $M$, and let $G = (Q,
Q\times Q)$ be the complete directed graph on $Q$. Given $q_1, q_2 \in
Q$, $simple(q_2,q_1)$ is the set of finite simple paths in $G$ from
$q_2$ to $q_1$. Since $G$ is finite, $simple(q_2,q_1)$ is also
finite. Given a finite path $\beta\in Q^\ast$, $edges(\beta)$ is the
set of edges in $\beta$.  Given a set of edges $W\subseteq
(Q\times Q)$,  $IO(W)\subseteq Q^\omega$ is the set of infinite paths
in which each edge in $W$ is visited i.o.

Let $V_\infty\subseteq Q$ be the set of states which have positive
probability of being visited i.o. under $f$, that is, for
each state $q$ in $V_\infty$, the set of paths in $Q^\omega$ that
visit $q$ i.o. has positive measure under $\mu_f$. Similarly, let
$E_\infty \subseteq V_\infty\times V_\infty$ be the set of edges that have positive
probability of being followed infinitely often, i.e., $E_\infty =
\set{e\in (Q\times Q) : \mu_f(IO(\set{e})) > 0}$. Let $G_\infty$ be the
directed graph $(V_\infty, E_\infty)$. We first show that each maximal
strongly connected component (MSCC) of
$G_\infty$ is also an ergodic set.

If $e = (q_1,q_2)$ is an edge in
$E_\infty$, then in order for an infinite path to to follow this edge i.o., it
must also travel from $q_2$ to $q_1$ i.o. Every finite path from $q_2$
to $q_1$ can be partitioned into a simple path from $q_2$ to $q_1$ and a finite
number of cycles. Thus for each $w\in IO(\set{e})$, there exists
$\beta \in simple(q_2,q_1)$, such that $w\in
IO(edges(\beta))$. Therefore $IO(\set{e}) \subseteq \bigcup_{\beta\in
  simple(q_2,q_1)} IO(edges(\beta))$. Since $\mu_f(IO(\set{e})) > 0$,
there exists at least one $\beta \in simple(q_2, q_1)$ such that
$\mu_f(IO(edges(\beta)) > 0$ and $edges(\beta)\in E_\infty$. 
Thus each edge in $G_\infty$ can in effect be traversed in the
opposite direction by following some path in $G_\infty$. So $G_\infty$
does not have an MSCC with an outgoing edge, and thus, is a collection of ergodic sets. 

Next we show that there exists some ergodic set $X$ in $G_\infty$ such
that the highest parity in $X$ is odd. Given $q\in Q$, let
$A_q\subseteq Q^\omega$ denote the event that $q$ is the
highest parity state visited i.o. 
Since $f$ is winning, there
must be some $q\in Q$ such that $q$ has odd parity and the event $A_q$
 has positive probability. Then
 $q\in V_\infty$, and let $X\subseteq V_\infty$ be the ergodic set in $G_\infty$ that
 contains $q$. Let $B_q\subseteq Q^\omega$ be the set of paths that
 visit $q$ i.o. and leave $X$ at most finitely many times.
Since, by the definition of $G_\infty$, it is not
 possible for a path to leave $X$ i.o. with positive probability, we
 get $\mu_f(A_q - B_q) = 0$, and therefore $\mu_f(A_q) = \mu_f(A_q\cap
 B_q)$.
 Now the probability that a suffix of a path remains in $X$,
 but does not visit some $q'\in X$ is zero. This is because, $X$ is
 strongly connected, and so avoiding $q'$ loses a positive
 amount of probability infinitely many times. In the limit, the
 probability of remaining in $X$ and never visiting $q'$ goes to zero.
If there is some $p\in X$
 such that the parity of $p$ is greater than the parity of $q$, then
 all paths in $A_q\cap B_q$ must have suffixes that avoid $p$, and so
 $\mu_f(A_q\cap B_q) = 0$, which
 contradicts that $A_q$ has positive probability. Therefore $q$ has
 the highest parity in $X$.

Finally, since each state in $X$ is visited i.o. with positive
probability, then the probability of visiting some state in $X$
starting from the start state $q_0$ must be positive. Let $\pi\in
Q^\ast$ be the shortest finite path starting from $q_0$ and ending in
$X$, such that $\mu_f(\pi\cdot Q^\omega) > 0$. 

We now define a memoryless strategy $g: Q\rightarrow Dist(\Sigma_I)$
that is winning for the environment. We first consider the case when $q\in V_\infty$.
Let $succ(q) = \{q': \exists (q,q')\in E_\infty\}$ be the successors
of $q$ in $G_\infty$. Given $a\in \Sigma_I$, we define $N_q(a) =
\set{q'\in Q: \delta(q,a)(q') > 0}$, and $D_q = \set{b\in \Sigma_I :
  N_q(b) \subseteq succ(q)}$. Given $p\in Q$ and $\beta\in Q^\ast$, we
say that $p$ is \emph{activated} by $f$ at
$\beta\cdot q$, if $\Sigma_{a\in \Sigma_I}f(\beta\cdot q)(a)\,
\delta(q,a)(p) > 0$. 
 If $D_q$ is empty, then this implies that, for all $\beta\in Q^\ast$,
whenever some $q'\in succ(q)$ is activated by $f$ at $\beta\cdot q$, some
$q''\not\in V_\infty$ must also be activated by $f$ at $\beta\cdot
q$. Then any time a path visits $q$, there is a positive probability
of visiting a state in $Q-V_\infty$ next. So a path that visits $q$ and
remains in $V_\infty$ loses some finite amount of probability. In the
limit, a path visiting $q$ i.o. must have probability zero because any
such path has a suffix in $V_\infty^\omega$. This contradicts $q\in
V_\infty$. Thus $D_q$ is non-empty for all $q\in V_\infty$.
We define $g: V_\infty\rightarrow Dist(\Sigma_I)$ as follows: for
$q\in V_\infty$, $g(q)$ is distributed uniformly over $D_q$ and is $0$
elsewhere. We extend $g$ to all of $Q$ as follows: for states in
$\pi$, we chose the value of $g$ such that edges in $\pi$ have
positive probability under $\mu_g$, and for all other states we let
$g$ take an arbitrary value. Then $g$ is a memoryless strategy since it
is a function $Q\rightarrow Dist(\Sigma_I)$. Consider the
graph $G_g$ induced by $g$ on $Q$. Every edge in $E_\infty$ is also an edge
in $G_g$, and no edges that leave $V_\infty$ have been
added. Also, all edges in $\pi$ are also in $G_g$. So the set $X\subseteq
V_\infty$ is a reachable ergodic set of $g$. Since the highest parity in
$X$ is odd, $g$ is a winning strategy.
\qed

% Given a
% memoryless strategy $f$, and a state $q\in Q$, we define the
% \emph{support} of $f$ at $q$, denoted $support(f(q))$,  as the set
% of inputs that have positive probability of being chosen by $f$ at
% $q$. 
%$Formally \{q'\in Q: \exists a\in\Sigma_I, f(q)(a) > 0 \text{ and }
% \delta(q,a)(q') > 0\}$. 

\begin{lem}\label{lemma:pure-strategy}
Let $M$ be a transducer and $f$ be a winning memoryless strategy for the
environment. Then there exists a memoryless and pure strategy $g$ such that
$g$ is winning.
\end{lem}
\proof 
Let $M = (\Sigma_I,\Sigma_O, Q, q_0, \delta, F, L)$. 
Given two memoryless strategies $f$ and $g$, we say that $g$
\emph{refines} $f$, iff $\forall q\in Q$, $\forall a\in \Sigma_I$,
$g(q)(a) > 0$ implies $f(q)(a) > 0$. The set of inputs chosen with
positive probability at state $q$ by memoryless strategy $f$ is simply the
support of the distribution $f(q)$, denoted $support(f(q))$. Then $g$
refines $f$ iff  $\forall q\in Q$, $support(g(q)) \subseteq support(f(q))$.
Note that, if $g$ refines $f$,  then $G_g$ is a subgraph of $G_f$, and
each connected component of $G_g$ is contained in a connected component of $G_f$.

Now assume that $f$ is a winning memoryless strategy for the
environment. Since $f$ is winning, by Lemma \ref{lemma:odd-ergodic},
there must be at least one reachable ergodic
set $P\subseteq Q$ of $G_f$ such that the highest parity in $P$ is odd. Let
$q\in P$ be a state with the highest parity. Then if a memoryless
strategy $g$ refines $f$, such that $q$ lies in a reachable ergodic set of $G_g$, then
$g$ is also winning. This is because every ergodic set of $G_g$ that
contains $q$ must be contained within some connected component of $G_f$
containing $q$, and $P$ contains all such components. So the highest parity
in such an ergodic set of $G_g$ must also be odd.
Thus it suffices to give a procedure of stepwise refinement of $f$, keeping
$q$ in a reachable ergodic set at each step, that terminates in a pure
strategy. This is because, at each step of
the procedure, the refined strategy is winning, and so it is also
winning at the end. We detail a two stage procedure below. 

\vspace{4pt}
\begin{description}
\item[Stage 1] In the first stage we only modify $f$ for states within the
ergodic set $P$ and each state is only modified once. At each step we
maintain a set $S\subseteq P$ of previously selected
states. The modified strategy at step $k$ is denoted $f_k$. The set of
already selected states at step $k$ is denoted $S_k$. The procedure is
then defined inductively as follows:
\begin{enumerate}
\item $S_1 = \set{q}$, and $f_1$ agrees with $f$ on $Q -\set{q}$ and
  chooses some input $a\in support(f(q))$ deterministically at $q$.
\item $S_{k+1} = S_k \cup \set{p_k}$, where $p_k\in P - S_k$ is chosen
  such that there is an edge $(p_k,p_k')$ in  $G_{f_k}$ for some state $p_k' \in
  S_k$. $f_{k+1}$ agrees with $f_k$ on $Q - \set{p_k}$, and $f_{k+1}(p_k)$ chooses
input $a_k \in support(f_k(p_k))$ deterministically such that $\delta(p_k,a_k)(p_k') > 0$. 
\end{enumerate}
At each step, the size of $P-S$ decreases by one. The prodecure
terminates when $P-S$ is empty. This happens in $|P|$ steps. In order
to ensure that the inductive procedure is sound, we need to show that a
suitable choice for $p_k$ and $a_k$ exists at each step. 
We first prove that, for all $k< |P|$,  for all $q' \in Q-S_k$, all
edges leaving $q'$ in $G_f$ are also present in $G_{f_k}$. This is true
at the first step. If this is
true at step $k$, then it is also true at step $k+1$, since
$Q-S_{k+1}\subseteq Q- S_k$ and $f_{k+1}$ and $f_k$ have the same value on
states in $Q-S_{k+1}$, so no edges that leave states in
$Q-S_{k+1}$ are removed at step $k+1$. So the statement holds by
induction. Since $P$ is an ergodic set of $G_f$, for all $k< |P|$,
there is some edge $e_k$ in $G_f$ that starts in $P-S_k$ and ends
$S_k$.  Now, by the claim proven above, $e_k$ is also an edge in
$G_{f_k}$. Then the source vertex of $e_k$ can be chosen as $p_k$ in
step $k+1$. Also, because $e_k = (p_k, p_k')$ is an edge in $G_{f_k}$, there must be
some $b\in \Sigma_I$ such that $f_k(p_k)(b) > 0$ and
$\delta(p_k,b)(p_k') > 0$. Then we can choose $b$ as $a_k$. Therefore
the inductive construction is well defined.

Next we show that, for all $k\leq |P|$, $f_k$ refines $f$, and $q$ is
reachable in $G_{f_K}$ from every state in $S_k$.
Let $f_k$ refine $f$. Since $f_{k+1}$ and $f_k$ agree on states in $Q
- \set{p_k}$, and $support(f_{k+1}(p_k)) \subseteq support(f_k(p_k))$,
we have $f_{k+1}$ refines $f$.
 Let $q$ be reachable in $G_{f_k}$ from every state in
$S_k$. Since $S_{k+1} = S_k\cup \set{p_k}$, it suffices to show that
$q$ is reachable in $G_{f_{k+1}}$ from every vertex in $S_k$, and
there is an edge in $G_{f_{k+1}}$ from $p_k$ to some vertex in
$S_k$. The first part is true because $f_{k+1}$ and $f_k$ take the
same value on states in $Q_k$, and the second part follows directly
from the definition of $f_{k+1}(p_k)$. 

Let $f' = f_{|P|}$. Then $f'$ refines $f$, all edges leaving $Q-P$ in
$G_f$ are also edges in $G_{f'}$, and $q$ is reachable in $G_{f'}$ from all states
in $P$.
%At the end, $f$ has been modified to choose inputs
%deterministically at all states in $P$, and $q$ can be still be
%reached from every state in $P$ (but vice versa is not
%necessary). Further, the only edges that have been removed lie within
%$P$, and no edges have been added.

\vspace{4pt}
\item[Stage 2] Since $P$ is a reachable ergodic set of $G_f$, there
  exists a minimal path $\pi$ in $G_f$ that starts from $q_0$ and ends
  in some state in $P$.  Since the path is minimal, none of its edges
  lie in $P$. Then $\pi$ is also a path in $G_{f'}$. Let $\pi =
  q_0,q_1,\dotsc,q_n$ where $q_n\in P$. Then there exists $b_k\in \Sigma_I$ such
  that $f'(b_k) > 0$ and $\delta(q_k,b_k)(q_{k+1}) > 0$. We define a pure
  memoryless strategy $g$ as follows: for states in $P$, $q$ agrees with $f'$;
  for a state $q_k$ in $\pi$, $g$ chooses input $b_k$
  deterministically; and for a state $q'$ that is not in $P$ or $\pi$, $g$ chooses some
  input $b\in support(f'(q'))$ deterministically.

Then $g$ refines $f'$ by construction, and thus $g$ refines $f$.  In
order to prove that $g$ is also a winning strategy, it
suffices to show that $q$ belongs to a reachable ergodic set of $G_g$.

Now, by construction, $\pi$ is also a path in $G_g$, and so some state
in $P$ is reachable from the start state in $G_g$. Also, $q$ is reachable in $G_g$ from
all states in $P$. Therefore $q$ is reachable from the start state in
$G_g$. Since $P$ is an ergodic set of $G_f$, and $G_g$ is a subgraph of
$G_f$, therefore there is no path in $G_g$ from $q$ to a state in
$Q-P$. Therefore, if $p\in Q$ is reachable from $q$ in $G_g$, then $q$ is
also reachable from $p$ in $G_g$. Thus $q$ lies in a reachable ergodic set of $G_g$.
\qed
\end{description}

\noindent Memoryless strategies are important because they induce an ergodic
structure on the set of states. Ergodic sets are useful because they
enable us to replace probabilistic reasoning with combinatorial
reasoning. In particular, they have the following crucial properties:
(a) the suffix of a path is contained in
  some ergodic set with probability $1$, and (b) the suffix of a
  path is contained in a proper subset of an ergodic set with
  probability zero \cite{KS60}. 
%and \item  for each reachable ergodic set, the probability that a path visits it is positive.
This allows us to define the winning strategy condition in terms of graph reachability.

\begin{lem}\label{lemma:odd-ergodic}
Let $M$ be a probabilistic transducer and $f$ be a memoryless
strategy for $M$. Then $f$ is winning for the environment iff $G_{M,f}$ has a
reachable ergodic set whose highest priority is odd.
\end{lem}

\proof 
Let $Q$ be the set of states of $M$, $E\subseteq 2^Q$ be the set of
ergodic sets of $G_{M,f}$ and $X = \bigcup_{Y\in E} Y$ be the set of
all ergodic states. We use the following useful property of ergodic
sets \cite{KS60}: (a) \item the suffix of a path is contained in
  some ergodic set with probability $1$, and (b) the suffix of a
  path is contained in a proper subset of an ergodic set with
  probability zero. 
Formally, we have, for all $\beta\in Q^\ast$,
$\mu_f(\beta\cdot (Q-X)^\omega) = 0$, and for all $Y\in E$, $q\in Y$,
$\mu_f(\beta\cdot (Y-\set{q})^\omega) = 0$.  

Let $odd(Q^\omega)$ be the set of paths in $Q^\omega$ whose highest parity visited i.o. is odd. If the highest parity in each ergodic set is even, then every path in $odd(Q^\omega)$ must have a suffix that is either contained in $(Q-X)^\omega$ or is contained in $Z^\omega$, where $Z$ is a proper subset of some ergodic set. Thus $odd(Q^\omega)$ is contained in the union of $\bigcup_{\beta\in Q^\ast} \beta\cdot (Q-X)^\omega$ and $\bigcup_{\beta\in Q^\ast, Y\in E, q\in Y}\beta\cdot (Y-\set{q})^\omega$. The probability of both these sets of paths is zero under $\mu_f$. Thus $\mu_f(odd(Q^\omega)) = 0$, and $f$ is not winning for the environment.

Next, assume that there is a reachable ergodic set $Y'$ such that the
highest parity in $Y'$ is odd. Let $q'\in Y'$ be a state with this
parity. Since $Y'$ is reachable from the start state, there exists a
path $\pi\in Q^\ast$, such that $\pi$ starts from $q_0$ and ends in
$Y'$ and $\mu_f(\pi\cdot Q^\omega) > 0$. Since $Y'$ is an ergodic
set, the probability of a path leaving $Y'$ after reaching it is $0$
\cite{KS60}. So we also have $\mu_f(\pi\cdot Y'^\omega) > 0$. Consider the set of paths
$S = \pi\cdot Y'^\omega - \pi\cdot (Y' - \set{q'})^\omega$. Then each
path in $S$ visits $q'$ i.o., and therefore, $S\subseteq
odd(Q^\omega)$. Now  $\mu_f(\pi\cdot (Y'
- \set{q'})^\omega) = 0$, and therefore, $\mu_f(odd(Q^\omega)) \geq
\mu_f(S) = \mu_f(\pi\cdot Y'^\omega) > 0$. Thus, $f$ is winning for
the environment. 
\qed

When the underlying probabilistic transducer is a composition, ergodic
sets acquire additional structure. Given a composer $C$ and a memoryless
strategy $f$ for $\T_C$, if a reachable ergodic set $X$ of
$G_{\T_C,f}$ contains some state from a component $M$ of $\T_C$,  then
either $X$ is contained in $M$ or all the reachable states of $M$ are
contained in $X$. Formally: 

\begin{lem}\label{lemma:component-ergodic}
Let $C = (D, \L, \mathcal{M}, \M_0, \Delta, \lambda)$ be a composer
over $\L$ and $f$ be a memoryless strategy for $\T_C$. Let $\M_i\in
\mathcal{M}$ and $Q_i$ be the state space of $M_i$. Let $X$ be
a reachable ergodic set of $G_{\T_C, f}$ such that $X\cap (Q_i\times \set{i}) \neq \emptyset$.
Then either $X\subseteq Q_i\times \set{i}$ or $(Q_i\times \set{i})\cap
Y \subseteq X$, where $Y$  is the set of states of $\T_C$ that are
reachable under $f$.
\end{lem}
\proof 
%See the appendix.
Assume that $X\cap (Q_i\times \set{i}) \neq \emptyset$ and  $X$ is not
contained in $Q_i\times \set{i}$. Let $(q, i)\in X\cap (Q_i\times
\set{i})$ and $(q',j)\in X - (Q_i\times \set{i})$, for some $j\neq i$. 
Since $X$ is ergodic, there is a path $\pi$ in $G_{\T_C,f}$ 
from $(q',j)$ to $(q,i)$. Let $s$ be the first state along $\pi$
such that $s = (q'', i) \in Q_i\times \set{i}$. We claim that $q'' =  q_0^i$,
where $q_0^i$ is the start state of $M_i$. Let $s' = (q''', k)$, where
$k\neq i$, be the predecessor of $s$ in $\pi$.
By the definition of
$G_{\T_C,f}$,  there is an edge from $s'$ to $s$ only if $\T_C$ can
transition from $s'$ to $s$ on some input with positive
probability. By Definition \ref{def:composition}, $\T_C$ can transition from
$(q''', k)$ to $(q'', i)$ only if $q'''$ is a final state
of $M_k$ and $q''$ is the initial state of $M_i$. Thus $(q_0^i, i)$ is
in $X$.

Since $X$ is an ergodic set, if it
contains a state $s$ of $\T_C$, then it also contains all states reachable
under $f$ from $s$. By definition, every state in $(Q_i\times
\set{i})\cap Y$ is reachable under $f$ from $(q_0^i,i)$. Since $X$
contains $(q_0^i, i)$, it also contains all states in $(Q_i\times
\set{i})\cap Y$. 
\qed

Given a graph $G$, each of whose vertices is assigned a priority,
we say that $G$ has the \emph{odd ergodic property} if it has a reachable 
ergodic set whose highest priority is odd.
Consider a composer $C$ and a memoryless strategy $f$ for
$\T_C$. Then, by Lemma \ref{lemma:odd-ergodic}, $f$ is winning for the
environment iff $G_{\T_C,f}$ has the odd ergodic property.
So the probabilistic notion of winning strategy is reduced to a
combinatorial one. However, the graph $G_{\T_C,f}$ is very large as it
contains all the internal states of each component
explicitly. Further, to show that $C$ satisfies $\alpha$, we have to
consider every possible memoryless strategy for $C$.
We tackle this complexity by simplifying the description of a strategy
$f$ and graph 
$G_{\T_C,f}$ so as to abstract away the
inner states of components and the choices that $f$ makes on those
inner states. Let $\mathcal{M}$ be the state space of $C$. We aim to
replace $G_{\T_C,f}$ by a simpler graph $G'$, whose set of vertices 
is $\mathcal{M}$, such that the odd ergodic property is preserved. We
first discuss this transformation informally, and 
then give formal definitions and proofs.

Let $M$ be a component of $\T_C$. If some reachable ergodic set of $G_{\T_C,f}$
lies entirely within $M$, we say $M$ is a \emph{sink}. When the
highest priority in the ergodic set is odd (resp. even) we say $M$ is
an \emph{odd} (resp. \emph{even}) sink for $f$. Note that a component
can be both an odd and an even sink for a given strategy.
Intuitively, we aim to replace the subgraph of
$G_{\T_C,f}$ that corresponds to states of $M$ by a single new 
vertex $x_M$ to obtain a new graph $G'$ and assign a suitable priority
to $x_M$ such that the odd ergodic property is preserved by the
transformation. Now if $M$ is not a sink, then, by Lemma
\ref{lemma:component-ergodic}, $x_M$ lies in a reachable ergodic set
of $G'$ iff all reachable states of $M$ lie in a reachable ergodic set
of $G_{\T_C,f}$. In this case, we can simply assign the highest
reachable priority in $M$ to $x_M$ and the odd ergodic property is
preserved. If, however, $M$ is a sink, then 
the collapse of $M$ to a single vertex might introduce new ergodic
sets in the graph. That is, $x_M$ might lie in an ergodic set of $G'$
which has no analogue in $G_{\T_C,f}$. We then have to choose the priority
of $x_M$ such that the odd ergodic property is still preserved. There are
two cases to consider:

\begin{itemize}
%\item $M$ does not contain any ergodic states. 
 % Then the priorities visited inside $M$ do not matter and we
 % can assign an arbitrary priority to $x_M$. 
%\item $M$ does contain an ergodic state belonging to a reachable
  %ergodic set $X$ of $G_{\T_C,f}$, but is not a sink. 
  %Then, by Lemma \ref{lemma:component-ergodic}, all reachable states of
  %$M$ lie in$X$. Then the new vertex $x_M$ will also lie in some
  %reachable ergodic set $X'$ of $G'$ and further $X'$ and $X$ will
  %have the same highest priority if we assign the highest reachable
  %parity in $M$ to $x_M$. 
\item $M$ is an odd sink for $f$. 
  Then, by Lemma \ref{lemma:odd-ergodic}, $f$ is winning for the
  environment. Let $f_M$ denote $f$ restricted to the states in
  $M$. Then $f_M$ is a memoryless strategy for $M$ that is winning for
  the environment, and in every composition involving $M$, the
  environment can simply play $f_M$ on the states in $M$ to win. So
  a component that is an odd sink is not useful for synthesizing
  compositions. We note that it is easy to check
  for and remove any odd sinks from $\L$ in a preprocessing step before
  attempting synthesis. Checking whether a particular component is a
  sink is equivalent to model checking Markov decision processes and can
  be done in polynomial time \cite{Var85b}. In the rest of the paper, we assume
  that the given library $\L$ does not contain components that are odd
  sinks.

\vspace{4pt}
\item $M$ is an even sink for $f$ but not an odd sink for $f$.
  Then, by Lemma
  \ref{lemma:component-ergodic}, every reachable state in $M$ either
  lies in an even sink or does not lie in an ergodic set. So no
  reachable state in $M$ is part of an ergodic set with odd highest
  priority. Thus collapsing $M$ to $x_M$ 
  does not remove any ergodic sets with odd highest priority. It only
  remains to consider the possibility that the transformation can
  introduce a new ergodic set whose highest priority is odd. We can
  avoid this by assigning
  a priority of $2\max(\alpha)$ to $x_M$, where $\max(\alpha)$ is the
  highest parity assigned by the index function $\alpha$. Then if
  $x_M$ is part of a reachable ergodic set $X'$ in $G'$, then $X'$ has
  highest priority $2\max(\alpha)$, which is even. Thus the odd ergodic
  property is preserved.
% Consider the vertex $x_M$ in $G'$. If $x_M$ is not part of a reachable
%  ergodic set  of $G'$, then the priority assigned to $x_M$ does not
%  matter. If $x_M$ is in a reachable ergodic set $X'$ of $G'$, then
%  since $2\max(\alpha)$ is  the highest possible priority by
%  construction, the highest priority  of $X'$ is even. 
\end{itemize}
%\vspace{4pt}

\noindent In formalizing the approach given above, instead of explicitly
transforming $G_{\T_C,f}$ into a more abstract graph, it is simpler to
directly define a suitable graph on the state space
$\mathcal{M}$ of the composer $C$ such that the odd ergodic property
is preserved. Just as a memoryless strategy $f$ applied to the
composition $\T_C$
gives rise to the graph $G_{\T_C,f}$, we define a combinatorial object,
called a \emph{choice function}, such that choice function $g$
together with composer $C$ gives rise to a graph $G_{C,g}$. 

%\vspace{4pt}
\begin{defi}[Choice Function]\label{def:choice}
Given a library $\L$ with width $D$ and index function $\alpha$, we
define the set $LABELS(\L)\subseteq 2^D\times \{1,\dotsc,
2\max{(\alpha)}\}\times \L$ as 
follows: $(X, j, M) \in LABELS(\L)$ iff there exists a memoryless strategy
$f$ for $M$ such that  
\begin{itemize}
\item $X\subseteq D$ is the set of exits of selected by $f$ in $M$.
\item If $M$ is an even sink for $f$, then $j = 2\max(\alpha)$.
\item Otherwise $j$ is the highest priority visited by $f$ in $M$.
\end{itemize}\medskip

\noindent Given a composer $C = (D,\L,\mathcal{M}, \M_0,\Delta, \lambda)$ over
$\L$,  a \emph{choice function} for $C$, is a function
$g:\mathcal{M}\rightarrow 2^D\times\{1,\dotsc, 2\max{(\alpha)}\}$, such
that, for all $\M_i\in \mathcal{M}$, $(g(\M_i), M_i) \in
LABELS(\L)$. The graph induced by $g$ on $C$, denoted $G_{C,g}$, is
the directed graph $(\mathcal{M}, \mathcal{E})$, where $(\M_1,\M_2)\in 
\mathcal{E}$ if $\Delta(\M_1, i) = \M_2$ for some $i\in D$ such that
$i\in X$ where $g(\M_1) = (X,j)$. The priority of a vertex $\M\in
\mathcal{M}$ of $G_{C,g}$ is $j$ where $g(\M) = (X, j)$.
We say that $g$ has \emph{rank} $r$, if $G_{C,g}$ has a
reachable ergodic set whose highest priority is $r$. 
\end{defi}

The size of the set $LABELS(\L)$ is at most $\max(\alpha) |\L| 2^{|D|}$. 
For an arbitrary triple $(X, j ,M)$, we can check whether $(X,j,M)\in
LABELS(\L)$  in time polynomial in $|M|$ using standard techniques for solving
Markov decision processes \cite{Var85b}.
% The number of distinct memoryless and pure strategies for a given $M \in \L$ is
% $|Q|^{|\Sigma_I|}$, where $Q$ is the set of states of $M$. Given a
% particular memoryless strategy $f$, the values of $X$ and
% $j$ can be calculated in time polynomial in $|Q|$ by using standard
% techniques for solving Markov chains \cite{Put94}. 
Thus $LABELS(\L)$ can be computed in time exponential in the size of $\L$.

%Specifying a  memoryless strategy for a composition is equivalent to 
%specifying a memoryless strategy
%for each of its components. This is a simple consequence of the fact
%that the states of a 
%composition can be partitioned into the states of its components.
%We use this equivalence to describe a memoryless strategy for a composition
%as a set of memoryless strategies for its components. We first look at how
%different memoryless strategies for a particular component can affect
%the overall strategy. Note that for the overall strategy to win, some
%ergodic set of $G_{\T_C,f}$ must have odd highest priority. A component
%strategy can affect this in only two ways: changing the priorities
%visited within the component, and changing the exits selected. If the
%component has no ergodic states, then the 
%priorities visited do not matter at all. 

\begin{theorem}\label{theorem:rank}
Let $C$ be a composer over $\L$. Then there exists a strategy for
$\T_C$ that is winning for the environment iff there exists a choice
function for $C$ that has an odd rank. 
\end{theorem}
\proof 
Let $C = (D, \L, \mathcal{M}, \M_0, \Delta, \lambda)$. Let $Q_i$ be
the state space of $M_i = \lambda(\M_i)$, for $\M_i\in \mcal$, 
and let $Q =  \bigcup (Q_i\times\{i\})$ be the state space of $\T_C$. 

\vspace{4pt}
\begin{description}
\item[\emph{Only If}] Assume there exists a strategy for $\T_C$
that is winning for the 
environment. Then, by Theorem \ref{theorem:memoryless}, there exists a
memoryless winning strategy $f$. We construct a choice function $g$
for $C$ as follows: for all $\M_i\in\mcal$, $g(\M_i) = (X,p)$,
where $X$ is the set of exits of $M_i$ selected by $f$, and $p =
2\max(\alpha)$ if $M_i$ is an even sink for $f$ and otherwise $p$ is the 
highest priority in $M_i$ visited by $f$. 
Since $f$ is winning, $G_{\T_C, f}$ 
has a reachable ergodic set $H$ with odd highest priority $r$. 
Consider the set $\mathcal{H}\subseteq \mcal$ defined as follows:
for all $\M_i\in\mcal$, $\M_i\in\mathcal{H}$ if $(Q_i\times \{i\})
\cap H \neq \emptyset$. Thus, $\mathcal{H}$ contains a state of the
composer $C$ if the corresponding component of $\T_C$ overlaps with
the ergodic set $H$. Since $\L$ contains no components that are odd
sinks, and even sinks can not be a part of an ergodic set whose
highest priority is odd, $H$ must contain
all the reachable states in each component named in $\mathcal{H}$. 

We claim that $\mathcal{H}$ is an ergodic set of $G_{C,g}$. We first
show that $\mathcal{H}$ is strongly connected. Let $\M_i$
and $\M_k$ be in $\mathcal{H}$. Since all the reachable states of
$M_i$ and $M_k$ are contained in $H$, in particular their start states
are also contained in $H$. Let these be $q_i$ and $q_k$
respectively. Then there is a path in $G_{\T_C,f}$ from $(q_i,i)$ to
$(q_k,k)$ because $H$ is an ergodic set of $G_{\T_C,f}$. Consider the
path $\pi$ from $(q_i,i)$ to $(q_k,k)$ that contains the least number of
exit states. Let the length of $\pi$ be $n$ and let $(q'_i,i)$ be
the first exit state along $\pi$. Suppose $\Delta(\M_i, x) = \M_j$,
where $q'_i$ is the exit state of $M_i$ in direction $x$, and let $q_j$
be the start state of $M_j$. Then, if $g(\M_i) = (X, p)$, we have
$x\in X$, so there is an edge from $\M_i$ to $\M_j$ in $G_{C,g}$, and
the immediate next state after $(q'_i,i)$ in $\pi$ is $(q_j,j)$. The
suffix of $\pi$ starting from $(q_j,j)$ is a path $\pi'$ from
$(q_j,j)$ to $(q_k,k)$ of length less than $n$. 
Further, by construction, among all such paths
it has the least number of exit states. Assume, by the induction
hypothesis, there is a path from $\M_j$ to $\M_k$ in $G_{C,g}$. Since
$(\M_i,\M_j)$ is also an edge in $G_{C,g}$, therefore, by induction,
there is a path 
from $\M_i$ to $\M_k$ in $G_{C,g}$. $\M_i$ and $\M_k$ were chosen
arbitrarily in $\mathcal{H}$. So $\mathcal{H}$ is strongly
connected. 

Next, we show that there are no edges that leave $\mathcal{H}$. 
Assume there is some edge in $G_{C,g}$ from a vertex
$\M_i\in \mathcal{H}$ to a vertex $\M_j \in \mcal -\mathcal{H}$. Let
$g(\M_i) = (X, p')$. Then there exists $x\in X$ such that
$\Delta(\M_i, x) = \M_j$. Let $(q', i)$ be the exit state of $M_i$ in
direction $x$. Then $(q', i)$ is reachable under $f$ and so is $(q_j,
j)$, where $q_j$ is the start state of $\M_j$. Therefore, there is an
edge in $G_{\T_C,f}$ from $(q',i)\in H$ to $(q_j,j)\not\in H$, which
contradicts that $H$ is an ergodic set. Thus no edges leave
$\mathcal{H}$ in $G_{C,g}$ and $\mathcal{H}$ is
ergodic. 

Finally, we show that the highest priority in $\mathcal{H}$ is
$r$. By construction of $g$, since $H$ does not contain any even
sinks, the priority of a vertex $\M_i$
in $\mathcal{H}$ is the highest priority visited in $M_i$ by
$f$. Thus, the highest priority in $\mathcal{H}$ is 
at most the highest priority in $H$, which is $r$. 
Let $(q,j) \in H$ be such that $q$ has priority $r$. Then the
highest priority visited by $f$ in $M_j$ is $r$, so $g(\M_j) = (X,r)$ for
some $X\subseteq D$. Since $\M_j\in\mathcal{H}$, the highest priority in
$\mathcal{H}$ is $r$, and $g$ has rank $r$.

\vspace{4pt}
\item[\emph{If}] Now assume that $g$ is a choice function for $C$ with
rank $p$, for some odd $p\leq \max(\alpha)$.
Then, by the definition of choice function, for all $\M_i\in\mcal$,
there exists a 
memoryless strategy $f_i$ for $M_i$, such that $g(\M_i) = (X_i, p_i)$
where $X_i$ is the set of exit directions of $M_i$ under $f_i$, and
$p_i = 2\max(\alpha)$ if $M_i$ is an even sink for $f_i$ and otherwise
$p_i$ is the highest priority visited by $f_i$. 

We define a memoryless strategy $f$ for $\T_C$ as follows:
for all $q\in Q_i$, $f(q,i) = f_i(q)$.  Since $g$ has rank $p$, there
exists a reachable ergodic set $\mathcal{H}\subseteq \mcal$ of $G_{C,g}$ with
highest priority $p$. 
Consider the set $H = \{(q,i) : q \in Q_i, \M_i\in
\mathcal{H}\}$, which consists of all states in all components
corresponding to the set $\mathcal{H}$. Let $H_f$ be the subset of $H$
that is reachable under $f$ from the start state of $\T_C$. We first
show that $H_f$ is strongly connected. Let $(q_i, i)$ and $(q_k, k)$ be
two arbitrary states in
$H_f$. Then $q_i$ is a state of $M_i$ and $q_k$ is a state of
$M_k$. Further, $\M_i$ and $\M_k$ are both in $\mathcal{H}$. We have the
following two cases: 

\begin{enumerate}
\item \emph{$q_i$ is the start state of
$M_i$}. Consider the shortest path in $G_{C,g}$ from $\M_i$ to
$\M_k$. Such a path exists because $\mathcal{H}$ is an ergodic set of
  $G_{C,g}$. Let the length of the path be $n$ and let $\M_j$ be the 
successor of $\M_i$ in this path. So there is path of length $n-1$ in
$G_{C,g}$ from $\M_j$ to $\M_k$. Now, by the definition of $G_{C,g}$,
there exists $x\in D$ such that $\Delta(\M_i, x) = \M_j$ and the exit
state in direction $x$ is reachable from the start state of
$\M_i$ under $f_i$. Thus there is a path in $G_{\T_C, f}$ from $(q_i, i)$ to $(q_j, j)$
where $q_j$ is the start state of $M_j$. By induction, there is a path
in $G_{\T_C, f}$ from $(q_i,i)$ to $(q_k,k)$.

\item \emph{$q_i$ is not the start state of $M_i$}. Let $g(\M_i) = (X, p')$,
where $X\subseteq D$. Since $p$ is the highest priority in $\mathcal{H}$
and $\M_i \in \mathcal{H}$, we have $p'\leq p \leq \max(\alpha)$. Thus
$p'\neq 2\max(\alpha)$ and so $M_i$ is not an even sink for $f$. 
Also, the
library $\L$ is assumed to have no components that are odd
sinks. Thus, some exit of $M_i$ 
must be reachable from $q_i$ under $f_i$. Let this exit be in
direction $x\in D$, and let $\Delta(\M_i, x) = \M_j$. 
Then there is a path in $G_{\T_C, f}$
from $(q_i, i)$ to $(q_j, j)$ where $q_j$ is the start state of
$M_j$. Now, since $q_j$ is a start state, by the previous case, there
is a path from $(q_j,j)$ to $(q_k,k)$ in $G_{\T_C,f}$. So there is a
path from $(q_i,i)$ to $(q_k,k)$ and therefore $H_f$ is strongly
connected.
\end{enumerate}

\noindent Assume that some edge in $G_{\T_C,f}$ leaves $H_f$. Let there be an
edge between $(q,i)\in H_f$ and $(q',j)\in Q - H_f$. Now $\M_j$ can
not belong to $\mathcal{H}$ because otherwise $(q', j)$ would be in
$H_f$. So we have $i\neq j$ and $(q,i)$ must be an exit state of
$M_i$. Therefore there is an edge in $G_{C,g}$ from $\M_i\in
\mathcal{H}$ to $\M_j\in \mcal - \mathcal{H}$, which contradicts that
$\mathcal{H}$ is ergodic. Thus $H_f$ is also an ergodic set.

By Lemma \ref{lemma:odd-ergodic}, it suffices to show that the highest
priority in $H_f$ is odd. Now $p$ is the
highest priority in $\mathcal{H}$, and $p$ is odd, which means $p\neq
2\max(\alpha)$. So there must exist $\M_i\in
\mathcal{H}$ such that some state $q$ in $M_i$ has priority $p$ and is
reachable under $f_i$. Then $(q, i)$ is in $H_f$ and so $H_f$ has 
highest priority at least $p$. Assume some state $(q', j)$ in $H_f$ has
priority $p' > p$. Since $q'$ is reachable under $f_j$, therefore, we have
$g(\M_j)  = (X, p'')$, for some $X\subseteq D$ and $p'' \geq p' >
p$. This contradicts the fact that $\M_j\in \mathcal{H}$. Thus the
highest priority in the ergodic set $H_f$ is $p$, which is odd.\qed
\end{description}

\noindent Let $\Gamma = LABELS(\L)$.  A composer and choice function pair has a
natural representation  as a regular $\Gamma$-labeled
$D$-tree.  Given a composer $C = (D,\L,\mathcal{M}, \M_0,\Delta,
\lambda)$ over $\L$, and a choice function $g$ for $C$, we denote by
$tree(C,g)$, the regular $\Gamma$-labeled full $D$-tree $\tuple{D^\ast,
\tau}$, where for all $x\in D^\ast$, we have that $\tau(x) =
(g(\Delta^\ast(x)),\lambda(\Delta^\ast(x)))$. Thus $tree(C,g)$ is the
tree obtained as a result of adding labels to $tree(C)$ such that a
node $x$ corresponding to $\M_i\in \mathcal{M}$ that is labeled with
$M_i$ in $tree(C)$ is labeled with $(X,j, M_i)$ where
$(X,j) = g(\M_i)$. 
As we show in the next lemma, the mapping is reversible, in the sense that given a
regular $\Gamma$-labeled $D$-tree, we can obtain a composer and choice
function in a natural way.

\begin{lem}\label{lemma:choice-tree}
Let $T$ be a regular $\Gamma$-labeled full $D$-tree. Then there exist
a composer $C$ over $\L$ and a choice function $g$ for $C$ such that
$tree(C,g) = T$.
\end{lem}

\proof 
Since $T$ is regular, there exists a deterministic transducer
$A = (D, \Gamma, Q, q_0, \delta, \lambda)$ that generates $T$. We define $C =
(D, \L, \mathcal{M}, \M_{q_0}, \delta', \lambda')$ as follows: for all
$q\in Q$,
\begin{itemize}
\item there is a state $\M_q$ in $\mathcal{M}$ 
\item if $\lambda(q) = (X,j, M_i)$ then $\lambda'(\M_q) = M_i$
\item for all $x\in D$, $\delta'(\M_q,x) = \M_{q'}$ where $q' = \delta(q,x)$
\end{itemize}\medskip

\noindent We define $g: \mathcal{M} \rightarrow 2^D\times \{1,\dotsc, k\}$ as
follows: for all $q\in Q$, $g(\M_q) = (X,j)$ where $\lambda(q) =
(X,j,M_i)$. Then, since $(X,j,M_i)\in \Gamma = LABELS(\L)$, $g$ is a
choice function.

Let $T = \tuple{D^\ast,\tau_1}$ and $tree(C,g) = \tuple{D^\ast,\tau_2}$. 
We need to show that $\tau_1 = \tau_2$. Consider a node $x\in
D^\ast$. We have $\tau_1(x) = \lambda(\delta^\ast(x))$ and $\tau_2(x) =
(g(\delta'^\ast(x)),\lambda'(\delta'^\ast(x)))$. Let $\delta^\ast(x) =
q$ and $\lambda(q) = (X,j,M)$. Then, by construction
of $C$ and $g$, $\delta'^\ast(x) = \M_q$, $g(\delta'^\ast(x)) =
g(\M_q) = (X,j)$, and $\lambda'(\delta'^\ast(x)) = \lambda'(\M_q) =
M$. Therefore $\tau_2(x) = (X,j, M) = \tau_1(x)$.
\qed

In light of Lemma \ref{lemma:choice-tree}, we can represent an
arbitrary regular $\Gamma$-labeled full $D$-tree as $tree(C,g)$ for some composer $C$
over $\L$ and some choice function $g$ for $C$. Similarly, we can
represent an arbitrary regular $\L$-labeled full $D$-tree as $tree(C)$ for some
composer $C$ over $\L$.

Since the question of whether a given composition satisfies $\alpha$
boils down to whether its composer has a choice function that has an
odd rank, we find it useful to characterize regular trees that correspond
to choice functions having a particular rank (see \cite{Sch06} for
related results). First, we inductively 
define the set of \emph{marked} nodes of a $\Gamma$-labeled $D$-tree
 as follows: the root is always marked, and
a node $y\cdot i$, where $i\in D$ and $y\in D^\ast$, is marked
if $y$ is marked and $i\in X$, where $(X,j,M)$ is the label on $y\cdot
i$.

\begin{lem}\label{lemma:rank}
Let $C = (D, \L, \mathcal{M}, \M_0, \Delta, \lambda)$ be a composer
over library $\L$ with width $D$, $\alpha$ be an index function for
$\L$, $g$ be a choice function for $C$, and $p\leq
\max{(\alpha)}$. Then $g$ has rank $p$ iff $tree(C,g)$ has a full
subtree $T$ such that: 
\begin{enumerate}
\item The root of $T$ is marked.
\item Every node in $T$ that is marked has priority label at most
  $p$.
\item From each marked node in $T$ there is a path in $T$ to a marked
  node with priority label $p$.
\end{enumerate}
\end{lem}

\proof 
\emph{Only If:} Assume $g$ has rank $p$. Then, by definition, there
exists a reachable ergodic 
set of $G_{C,g}$ whose highest priority is $p$. Let $\M_i\in \mathcal{M}$
be a vertex of $G_{C,g}$ that lies in this ergodic set such that there
is a path in $G_{C,g}$ from $\M_0$
to $\M_i$ and $\M_i$ has priority $p$. Since $\M_i$ is
reachable from $\M_0$ in 
$G_{C,g}$,  there exists some $x\in D^\ast$ such
that $\Delta^\ast(x) = \M_i$ and $x$ is marked. Then the node
$x\in tree(C,g)$ is labeled 
with $(X, p, M_i)$ for some $X\subseteq D$. 
Let $T_x$ be the full subtree of $tree(C,g)$ rooted at $x$. We show that $T_x$ has the
desired property. Let $y$ be a node in $T_x$ that is marked
and let $\Delta^\ast(y) = \M_j$. Then $\M_j$ must lie in the ergodic set of
$G_{C,g}$ containing $\M_i$ and $g(\M_j) = (Y,p')$ for some
$Y\subseteq D$ and $p'\leq p$. So $y$ is labeled $(Y,p', M_j)$ and
has a priority label less than or equal to $p$. All that remains is to show that
some marked node in $T_x$ with a priority label $p$ is reachable from
$y$. Since $\M_i$ is reachable from $\M_j$ in 
$G_{C,g}$, there must exist $x'\in D^\ast$ such that
$\Delta^\ast(y\cdot x') = \M_i$ and $yx'$ is marked. 
Then $z = yx'$ is also labeled $(X, p, M_i)$. 
Since $T_x$ is a full subtree, and $y\in T_x$, therefore
$z$ also lies in $T_x$ and there is a path from $y$ to $z$.

\emph{If:} Let $T$ be a full subtree of $tree(C,g)$
that satisfies the given property. Consider the set $\H\subseteq
\mathcal{M}$ of vertices in $G_{C,g}$ defined as follows: $\M_i\in \H$
if there exists some marked node $x\in T$ such that $\Delta^\ast(x) =
\M_i$. Note that every vertex in $\H$ is reachable from $\M_0$ in
$G_{C,g}$ and has priority at most $p$.
Consider the subgraph $G_\H$ of $G_{C,g}$ induced by $\H$. Let $\H'$
be an ergodic set of $G_\H$ and let $\M$ be an arbitrary vertex in $\H'$. Then there exists a
marked node $y\in T$ such that $\Delta^\ast(y) = \M$. Let $z =
a_1a_2\dotsm a_n\in
D^\ast$ be such that $yz$ is marked and has priority label
$p$. Then every node along the path from $y$ to $yz$ is also
marked. Let $\M'_1 = \Delta^\ast(y)$ and $\M'_{i+1} =
\Delta^\ast(y a_1\dotsm a_i)$, for $1\leq i < n$. Then the
priority of $\M'_n$ is $p$ and $\M'_1,\M'_2,\dotsc,\M'_n$ is a path in
$G_\H$. Since $\M'_1\in \H'$ and $H'$ is an ergodic set of $G_\H$, $\M'_n$ must
also lie in $\H'$. Thus the highest priority in $\H'$ is $p$.

Finally, it suffices to show that no edges leave $\H$ in $G_{C,g}$, as this
implies that $\H'$ is also an ergodic set of $G_{C,g}$. Consider an
edge in $G_{C,g}$ from a vertex $\M\in \H$ to a vertex
$\M'\in \mcal$. Then there exist $X\subseteq D$ and $c\in X$ such that
$\Delta(\M,c) = \M'$ and $g(\M) = (X, j)$ for some priority $j$. Since
$\M$ lies in $\H$, there exists a marked  node $x\in T$ such that
$\Delta^\ast(x) = \M$. Then $x\cdot c$ is also marked and
$\Delta^\ast(x\cdot c) = \M'$. By the construction of $\H$, $\M'$ lies
in $\H$. Thus there are no edges that leave $\H$.
\qed

The conditions given by Lemma
\ref{lemma:rank} can be checked by a suitable tree automaton as follows:

\begin{lem}\label{lemma:automata}
Let $\L$ be a library with width $D$ and let $p\leq k$.  Then there
exists an nondeterministic B\"uchi tree automaton (NBT) $\A_p$ such that
$\A_p$ accepts a $\Gamma$-labeled regular $D$-tree $T$
iff $T = tree(C,g)$ for some composer $C$ over $\L$ and choice
function $g$ with rank $p$. 
\end{lem}

\proof 
By Lemma \ref{lemma:choice-tree} and \ref{lemma:rank}, it suffices to
construct an NBT $\A_p$ such that $\A_p$ accepts a tree $T'$ iff $T'$
has a full subtree $T$ that satisfies the three conditions in
Lemma \ref{lemma:rank}. For simplicity, the automaton is defined over binary
trees, where $D = \set{0,1}$, but the definition can be easily
extended to $n$-ary trees.

Let $\A_p = (\Gamma,Q, q_0, \delta,\beta)$. 
We define $Q = \{\mathsf{search}, \mathsf{cut}, \mathsf{wait},
\mathsf{reach}, \mathsf{visit}, \mathsf{err}\}$, $q_0=
\mathsf{search}$ and $\beta = \{\mathsf{visit},\mathsf{wait},\mathsf{cut}\}$.  
The states of the automaton can then be described as follows:

\vspace{4pt}
\begin{itemize}
\item $\mathsf{search}$: In this state the automaton is searching for the
  root of the special subtree.
\item $\mathsf{cut}$: This represents a branch not taken.
\item $\mathsf{wait}$ and $\mathsf{reach}$: In these states the
  automaton has entered the subtree and is looking for nodes labeled
  with $p$. 
\item $\mathsf{visit}$: In this state the automaton has
  just visited a node with label $p$ in the subtree.
\item $\mathsf{err}$: This is an error state that is entered if 
  there is a label higher than $p$ in the subtree. 
\end{itemize}

\vspace{4pt}
The transition function $\delta$ is defined as follows: For all $\rho
= (X, j, M_i)\in \Gamma$,
\begin{enumerate}
\item For $q\in \{\mathsf{cut}, \mathsf{err}\}$, $\delta(q, \rho) =
  \{(q, q)\}$. 

\item For $q = \mathsf{search}$
\begin{equation*}
 \delta(q,\rho) =  
\begin{cases}
 \{(\mathsf{search}, \mathsf{cut}), (\mathsf{wait}, \mathsf{cut})\}
 &\quad \text{if $X = \{0\}$} \\
 \{(\mathsf{cut}, \mathsf{search}), (\mathsf{cut}, \mathsf{wait})\}
 &\quad \text{if $X = \{1\}$} \\
 \{(\mathsf{search}, \mathsf{cut}), (\mathsf{cut}, \mathsf{search}),
 (\mathsf{wait}, \mathsf{wait})\} 
&\quad \text{if $X = \{0,1\}$} 
\end{cases}
\end{equation*}

\item For $q\in \{\mathsf{wait}, \mathsf{reach}, \mathsf{visit}\}$, if
  $j > p$ then $\delta(q,\rho) = \{(\mathsf{err},\mathsf{err})\}$, if
  $j = p$ then
\begin{equation*} 
%\text{if } j = p \text{ then } \;
\delta(q, \rho) = 
\begin{cases}
\{(\mathsf{visit}, \mathsf{cut})\} &\quad \text{if $X = \{0\}$} \\
\{(\mathsf{cut}, \mathsf{visit})\} & \quad \text{if $X = \{1\}$} \\
\{(\mathsf{visit}, \mathsf{visit})\} & \quad \text{if $X = \{0,1\}$} 
\end{cases}
\end{equation*}
and if $j <  p$ then
\begin{equation*}
%\text{and if } j < p \text{ then } \; 
\delta(q, \rho) = 
\begin{cases}
\{(\mathsf{reach}, \mathsf{cut})\} & \quad \text{if $X = \{0\}$} \\
\{(\mathsf{cut}, \mathsf{reach})\} & \quad \text{if $X = \{1\}$} \\
\{(\mathsf{reach}, \mathsf{wait}), (\mathsf{wait}, \mathsf{reach})\} & 
\quad \text{if $X = \{0,1\}$} 
\end{cases}
\end{equation*}
\end{enumerate}

\noindent In the first stage, $\A_p$ guesses the location of the
root of the special subtree $T$. While searching for this root, $\A_p$
remains in the state $\mathsf{search}$. When it encounters the root,
it enters the state $\mathsf{wait}$ for the first time. This starts
the second stage, where $\A_p$ considers only marked nodes in $T$. 
In directions that correspond to a
non-marked node, $\A_p$ moves to the state $\mathsf{cut}$ and remains
there perpetually. From every marked node in $T$, $\A_p$
guesses a path to another marked node with label
$p$, using the states $\mathsf{wait}$ and $\mathsf{reach}$. It starts
this search in state $\mathsf{wait}$, moves to state
$\mathsf{reach}$ immediately, remains there until it
encounters a marked node with label $p$, and then moves to state
$\mathsf{visit}$. If there is no path from some node to another node with
label $p$, all runs corresponding to the choice of $T$ as subtree will
eventually get stuck in $\mathsf{reach}$. Thus, some run corresponding to
$T$ as the required subtree is accepting iff $T$ satisfies
the required conditions. 
\qed

\begin{theorem}\label{theorem:embedded}
Let $\L$ be a library with width $D$, $R$ be an exit control relation
for $\L$, and $\alpha$ be an index function for $\L$. There exists a
non-deterministic parity tree automaton (NPT) $\B$ such that, for all
composers $C$ over $\L$, $\B$ accepts $tree(C)$ iff $C$ 
satisfies $\alpha$ and $C$ is compatible with $R$. Consequently, $\B$
is non-empty iff $\L$ realizes $\alpha$ under $R$.
\end{theorem}
\proof 
We define $\B = \B_R\cap \B_\alpha$, where $\B_R$ is a safety tree
automaton that accepts $tree(C)$ iff $C$ is compatible with $R$,
and $\B_\alpha$ is an NPT that accepts $tree(C)$ iff $C$ satisfies
$\alpha$. Since the intersection of a safety automaton and an NPT is
again an NPT, $\B$ is also an NPT.

\vspace{4pt}
\emph{Construction of $\B_R$}:
For simplicity, we define the automaton
for the case $D = \set{0,1}$, and note that the definition can be
easily extended for arbitrary $D$. $\B_R = \set{\L,
  \set{\mathsf{start}}\cup D, \mathsf{start}, \delta_R}$, where
  $\delta_R$ is defined as follows: For all $M\in \L$, 
\begin{itemize}
%\item For $q = \mathsf{fail}$, $\delta_R(q, M) = \set{(\mathsf{fail, fail})}$
%\item For $q\in D$, if $(q,M)\not\in R$ then $\delta_R(q, M) = \set{(\mathsf{fail,fail})}$
\item $\delta_R(\mathsf{start},M) = \set{(0, 1)}$
\item For $q\in D$, if $(q,M) \in R$ then $\delta_R(q,M) = \set{(0, 1)}$
\end{itemize}
Note that $\B_R$ has no transitions out of the states $0$ and $1$ iff
the exit control relation $R$ is violated. Thus $\B_R$ accepts
$tree(C)$ iff $C$ is compatible with $R$.

\vspace{4pt}
\emph{Construction of $\B_\alpha$}: 
Let $\Gamma = LABELS(\L)$ and
let $\A_p = (\Gamma,Q, q_0, \delta, \beta)$ be the NBT defined
in Lemma \ref{lemma:automata}. We define $\A'_p = (\L, Q, q_0, \delta',
\beta)$, where  \[\delta'(q, M_i)= \bigvee_{(X, j, M_i)\in LABELS(\L)}
\delta(q, (X, j, M_i))\]

While $\A_p$ accepts $\Gamma$-labeled $D$-trees, 
$\A'_p$ accepts $\L$-labeled $D$-trees.
$\A'_p$ simply simulates $\A_p$ by using its larger transition
function to guess the missing portion of the labels. We can
characterize the regular trees accepted by $\A'_p$ as 
follows: for a composer $C$ over $\L$, $\A'_p$ accepts
$tree(C)$ iff there exists a
choice function for $C$ which has rank $p$.

Consider the automaton $\A'_\alpha$ whose language is the union of the
language of each $\A'_p$, for all odd $p \leq \max(\alpha)$.
%Formally, $\A' = \bigcup_{\set{p\leq \max(\alpha) | p \text{ is odd}}}\A'_p$. 
Let $C$ be a composer over $\L$. Then $\A'_\alpha$ accepts
$tree(C)$ iff there exists a choice function for $C$ that
has an odd rank. Thus, by Theorem \ref{theorem:rank},  $\A'_\alpha$ accepts
$tree(C)$ iff $C$ does not satisfy $\alpha$. 
Finally, consider the automaton $\B_\alpha = \overline{\A'_\alpha}$,
which is the complement of $\A'_\alpha$. Then $\B_\alpha$ accepts
$tree(C)$ iff $C$ satisfies $\alpha$.

Since an NPT is nonempty iff it accepts a regular tree, and $\L$
realizes $\alpha$ under $R$ iff some composer $C$ over $\L$ satisfies
$\alpha$ and $C$ is compatible with $R$, therefore  $\B$ is non-empty
iff  $\L$ realizes $\alpha$ under $R$. 
\qed

The NBT $\A'_p$ accepts $|D|$-ary trees and has $O(1)$ states,
with an alphabet of size $|\L|$, so $\A'_\alpha$ is an NBT with $O(k)$ states,
where $k=\max(\alpha)$.  It follows that $\B_\alpha$ is a
nondeterministic parity  tree automaton (NPT) with $k^{O(k)}$ states
and parity index $O(k)$ \cite{MS95}. Also, $\B_R$ is a safety
automaton with $O(|D|)$ states. Thus, their intersection $\B$ is an NPT
with $|D|k^{O(k)}$ states and parity index $O(k)$, whose nonemptiness can be
tested in time $|\L||D|^{O(k+|D|)}k^{O(k^2+k|D|)}$ \cite{MS95}. 
We thus obtain the following:

\begin{theorem}
The embedded parity realizability problem is in EXPTIME. \qed
\end{theorem}

%myv1: added here also
If an alternating tree automaton is nonempty, then it must accept some
regular tree \cite{MS95}. Given a regular tree accepted by $\B$, we can
obtain a finite transducer that generates that tree. This transducer
is a composer that realizes $\alpha$ under $R$. Thus, we also obtain
a solution to the embedded parity synthesis problem.

\begin{theorem}
The embedded parity synthesis problem is in EXPTIME. \qed
\end{theorem}

The complexity of our solution is exponential in both~$k^2$, where~$k$
is the highest parity index, as well as~$|D|$, which is the number of
exit states in each component. 
%myv1:
The exponential dependence on $k$ is expected, as typical
algorithms for solving parity games are exponential in the parity
index, cf. \cite{EJS93, Sch07}.
%solving parity games already has a lower bound that is exponential in the 
%highest parity. 
%The exponential dependence on $k$ cannot be avoided as the complexity of 
%solving parity games already has a lower bound that is exponential in the 
%highest parity. 
Improving $k^2$ to $k$ is an open challenge.
It is also an open question whether the exponential dependence on~$|D|$ 
can be avoided.

%\noindent\colorbox{lightgray}{\parbox{\columnwidth}{
We remark that the embedded
parity synthesis problem can be viewed as a 2-player partial
information stochastic parity game. Informally, the game can be
described as follows: The two players are the composer C and the environment
E. The C player 
chooses components and the E player chooses paths through the
components chosen by C. C cannot see the moves E makes inside a
component. At the start C chooses a component $M$ from the library
$\L$. The turn 
passes to E, who chooses a sequence of inputs, inducing a path in $M$
from its start state to some exit $x$ in $D$. The turn then passes to C,
which must choose some component $M'$ in $L$ and pass the turn to E
and so on. As C
cannot see the moves made by E inside $M$, C cannot base its choice on
the run of E in $M$, but only on the exit induced by the inputs selected
by E and previous moves made by C. So C must choose the same next
component $M'$ for different runs that reach exit $x$ of $M$. In general,
different runs will visit different priorities inside $M$. This is a
two-player stochastic parity game where one of the players does not
have full information. If C has a winning strategy that requires a
finite amount of memory, then we can use such a strategy to obtain a
suitable finite composer that satisfies the index function $\alpha$,
thus solving the embedded parity synthesis problem. If C has no
winning strategy or if every winning strategy requires infinite
memory, then $\alpha$ is not realizable from the library $L$.

We also note that, when viewed in the framework of games, our result
is a rare positive result for partial-information stochastic
games. In general, 2-player partial information stochastic games are known
to be undecidable even for co-Buchi objectives (and thus for parity
objectives) \cite{CD10}.
%the best result one can hope for is to show that some
%restricted but useful class of partial information parity games is
%decidable. Our result on the embedded parity problem can be viewed as
%such a result. 

\section{Synthesis for DPW Specifications}
Let $A$ be a deterministic parity automaton (DPW), 
$M$ be a probabilistic transducer and $\L$ be a library of components. 
We say $A$ is a \emph{monitor} for $M$ (resp. $\L$) if the input alphabet 
of $A$ is the same as the output alphabet of $M$ (resp. $\L$).
Let $A$ be a monitor for $M$ and let $L_A$ be the language accepted by $A$.
We say a strategy $f$ for $M$ is \emph{winning} for the
environment iff $\mu_f(L_A) < 1$, i.e., the output of $M$ is rejected by 
$A$ with positive probability. We say that $M$ \emph{satisfies} $A$ if 
there exists no winning strategy for the environment.

\begin{defi}
The \emph{DPW probabilistic realizability problem} is:
Given a library $\L$ and a DPW $A$ that is a monitor for $\L$, decide
whether there exists a  composer $C$ over $\L$, such that $\T_C$
satisfies $A$. If such a composer exists, we say that $\L$
\emph{realizes} $A$. The \emph{DPW probabilistic synthesis problem} is to
find such a composer $C$ if it exists.
\end{defi}

We transform this problem into a version of the embedded parity
problem solved in Section \ref{sec:embedded-parity}.
Let $A = (\Sigma_O, Q_A, s_0, \delta_A, \alpha_A)$ be a DPW and $M =
(\Sigma_I,\Sigma_O, Q_M, q_0, \delta_M, F, L)$ be a probabilistic
transducer.
For $s\in Q_A$, we denote by $M\times A_s$, the probabilistic transducer 
$(\Sigma_I, \Sigma_O, Q_M\times Q_A, (q_0, s), \delta, F\times Q_A, L')$,
where $\delta((q,s'),a)(q',s'') = \delta_M(q,a)(q')$ if 
$s'' = \delta_A(s', L(q))$ and $0$ otherwise.
Given a library $\L$ with width $D$, we define the \emph{augmented library}
$\L_A = \{M\times A_s : M\in \L, s\in Q_A\}$. The width of $\L_A$ is
$D\times Q_A$. We define the exit control relation $R_A\subseteq
D\times Q_A\times \L_A$ for
$\L_A$ as follows: for all $i\in D$, $s\in Q_A$, $M\in \L$, we have
$(i,s,M\times A_s)\in R_A$. 
%$R_A = \set{(i,s,M\times A_s) : i\in D, s\in Q_A,M\in \L}$. 
We also extend $\alpha_A$ to $\L_A$ as
follows: for $(q,s')\in Q_M\times Q_A$, $\alpha_A(q,s') = \alpha_A(s')$. Thus
$\alpha_A$ is an index function for $\L_A$. 

Our first step is to treat this augmented library as a new library and 
solve the embedded parity synthesis problem for $\L_A$ with $\alpha_A$ as 
the index function and $R_A$ as the exit control relation.
This gives us a tree automaton that accepts
$\L_A$-labeled $(D\times Q_A)$-trees and that is empty iff $\L_A$ does not 
realize $\alpha_A$ under $R_A$. Later, we show how to transform
this automaton into another that accepts $\L$-labeled $D$-trees and is
empty iff $\L$ does not realize $A$. Since, by definition, $\L_A$ 
bijectively maps to $\L\times Q_A$, we find it convenient to use labels 
from $\L\times Q_A$ in place of $\L_A$. 
%We do this by slightly modifying the definition of composer to 
%work with $(\L\times Q_A)$-labeled trees instead of $\L_A$-labeled trees. 
We now define a composer for the augmented library. The states of
the composer are pairs of the form $(\M,s)$, where $s$ is a monitor
state and $\M$ represents an instance of a component from $\L$. 
%For such a state $(M,s)$, we call $M$ the \emph{library state} of the
%composer and $s$ the monitor state of the composer. 
A \emph{composer} for $\L_A$, is a
deterministic transducer $C = (D\times Q_A, \L\times Q_A, \mathcal{M}\times Q_A, (\M,s),
\Delta, \lambda)$. The following lemma follows directly from Theorem
\ref{theorem:embedded}\footnote{Note that even with the slightly
  modified definition of composer, the results of the 
  previous section still apply because a pair $(M,s)\in \L\times Q_A$
  still uniquely identifies an element of $\L_A$.}. 

\begin{lem}\label{lemma:embedded-DPW}
Let $\L$ be a library and $A$ be a DPW that
is a monitor for $\L$.
There exists an NPT $\mathcal{B}$ that  accepts a
regular tree $T$ iff $T = tree(C)$ for some composer $C$ over $\L_A$
such that $\T_C$ satisfies $\alpha_A$ and $C$ is compatible with $R_A$. \qed
\end{lem}

Given a composer $C$ over a library $\L$ and a monitor $A$ for $\L$,
we can extend $C$ to a composer over the augmented library $\L_A$. 
\begin{defi}[Augmented Composer]
Let $\L$ be a library and $A$ be a monitor for $\L$. Let $C = (D, \L, \mathcal{M}, \M_0, \Delta, \lambda)$ be a composer over $\L$. The \emph{augmentation} of $C$ by $A$, denoted $C_A$, is a composer over $\L_A$ such that $C_A = (D\times Q_A, \L\times Q_A, \mathcal{M}\times Q_A, (\M_0, s_0), \Delta', \lambda')$, where
\begin{itemize}
\item For all $s \in Q_A$, $\M\in \mathcal{M}$, $\lambda'(\M,s) = (\lambda(\M), s)$.
\item For all $i \in D$, $\M\in \mathcal{M}$ and $s,s'\in Q_A$, $\Delta((\M,s), (i,s')) = (\Delta(\M,i),s')$.
\end{itemize}
\end{defi}\medskip

\noindent We say $C_A$ is an augmented composer. While a composer only keeps 
track of the transfer of control between components, the augmented 
composer also keeps track of the state of the monitor before and after 
the control is transferred. 
To go from augmented composers to composers, we use techniques
from synthesis with incomplete information \cite{KV97c}.
We start by describing a relation between $tree(C)$ and $tree(C_A)$. 
First we need to introduce some convenient notation.

Let $X$, $Y$ and $Z$ be finite sets. For a $Z$-labeled $(X\times Y)$-tree 
$\tuple{T,V}$, we denote by $xray(Y, \tuple{T,V})$, the 
$(Z\times Y)$-labeled
$(X\times Y)$-tree $\tuple{T,V'}$ in which each node is labeled
by both its direction in $Y$ and its labeling in $\tuple{T,V}$.
We define operators  $hide_Y$ and
$wide_Y$. The operator $hide_Y : (X\times Y)^\ast
\rightarrow X^\ast$ replaces each letter $x\cdot y$, where $x\in X$
and $y\in Y$, by the letter
$x$. The operator $wide_Y$ maps $Z$-labeled $X$-trees to $Z$-labeled
$(X\times Y)$-trees as follows: $wide_Y(\tuple{X^\ast,V}) = \tuple{(X\times
Y)^\ast, V'}$, where for each node $w\in (X\times Y)^\ast$, we have
$V'(w) = V(hide_Y(w))$.

\begin{lem}\label{lemma:augmented}
Let $\L$ be a library and $A$ be a monitor for $\L$.
Let $C$ be a composer over $\L$ and $C_A$ be the augmentation of $C$ by 
$A$. Then $tree(C_A) = xray(Q_A , wide_{Q_A}(tree(C)))$.
\end{lem}

\proof 
Let $T$ be the unlabeled full $D$-tree and $T'$ be the unlabeled full
$(D\times Q_A)$-tree. 
Let $tree(C) = \tuple{T,V}$. Since $tree(C)$ is a $\L$-labeled $D$-tree, $wide_{Q_A}(tree(C))$ is a $\L$-labeled $(D\times Q_A)$-tree, and $xray(Q_A , wide_{Q_A}(tree(C)))$ is a $(\L\times Q_A)$-labeled $(D\times Q_A)$-tree. Let $xray(Q_A , wide_{Q_A}(tree(C))) = \tuple{T',V'}$. Now, by definition, $tree(C_A)$ is also a $(\L\times Q_A)$-labeled $(D\times Q_A)$-tree. Let $tree(C_A) = \tuple{T', V''}$. It suffices to prove that $V'' = V'$.

Let $C = (D, \L, \mathcal{M}, M_0, \Delta, \lambda)$ and $C_A = (D\times Q_A, \L\times Q_A, \mathcal{M}\times Q_A, (M_0, s_0), \Delta', \lambda')$. Let $w\in T'$ and let $(M,s)\in \L\times Q_A$ be the direction of $w$. Then $V'(w) = (V(hide_{Q_A}(w)), s) = (\lambda(M),s)$. Then $V''(u) = \lambda'(M,s) = (\lambda(M), s)$. Therefore $V'' = V'$.
\qed

\begin{theorem}\label{theorem:augmented}
Let $\L$ be a library and $A$ be a monitor for $\L$. 
Let $C$ be a composer over $\L$ and $C_A$ be the augmentation of $C$ by $A$. 
Then $C$ satisfies $A$ iff $C_A$ satisfies $\alpha_A$.
\end{theorem}

\proof
Let $A = (\Sigma_O, Q_A, s_0, \delta_A, \alpha_A)$ and $C = (D, \L,
\mathcal{M}, \M_0, \Delta, \lambda)$.
Let $Q$ and $Q'$ be the state spaces of $\T_C$ and $\T_{C_A}$,
respectively. Then $Q' = Q\times Q_A$. 
Let $q_0$ be the start state of $\T_C$. Then $(q_0,s_0)$ is the start
state of $\T_{C_A}$. Let $L_A$ be the language of $A$. Given $w\in
Q^\omega$, we denote by $out(w)$, the output sequence produced by
$\T_C$ corresponding to state sequence $w$. We define $L =
\set{w\in Q^\omega : out(w)\in L_A}$. Then a strategy $f$ for $\T_C$ is winning for the
environment iff $\mu_f(L) < 1$.

We define a notion of consistency for words in $Q'^\ast$ as
follows: $(q_0,s_0)$ is consistent, and if $\beta\in Q'^\ast$ is
consistent then, for all $q\in Q$, $\beta\cdot (q,\delta_A(s,q'))$ is
consistent, where $(q', s)$ is the last letter of $\beta$. An
infinite path in $Q'^\omega$ is \emph{consistent} if all of its finite
prefixes are consistent. We let $H$ denote the set of all  consistent paths in
$Q'^\omega$, and $T_H$ denote the subtree of $Q'^\ast$ that
contains all consistent words in $Q'^\ast$. Then $T_H$ contains all
paths in $H$. We define $R$ to be the set of paths in $Q'^\omega$
where the highest parity visited i.o. is even. 

Let $g$ be a strategy for $\T_{C_A}$ and
$\mu_g$ be the probability measure it induces on $Q'^\omega$. Then, by
the definition of $\L_A$, for every $\beta\in Q'^\ast$ that is not
consistent, we have $\mu_g(\beta\cdot Q'^\omega) = 0$. Therefore, the
probability that an infinite path over $Q'$ is not consistent is
zero. So consistent paths are the only ones that matter probabilistically.
In particular, given two strategies $g$ and $g'$ for $\T_{C_A}$, such
that $g(w) = g'(w)$ for all $w\in T_H$, we have $\mu_g = \mu_g'$. Thus,
in order to define a strategy for all of $Q'^\ast$ it suffices to
define it for $T_H$. Also, $g$ is winning for the environment iff
$\mu_g(H \cap R) < 1$, i.e., the probability that the highest parity
visted i.o. in a consistent path is positive.

Similarly, given a strategy $f$ over $\T_C$, we have $\mu_f(q_0\cdot
Q^\omega) = 1$, i.e., the probability of a path not beginning from the
start state is zero. This means that two strategies that agree on nodes in
$q_0\cdot Q^\ast$ induce the same distribution on
$Q^\omega$. Thus, in order to define a strategy for all of $Q^\ast$,
it suffices to define it for $q_0\cdot Q^\ast$.

Finally, we note that $T_H$ is isomorphic to $q_0\cdot Q^\omega$,
with the isomorphism $h:T_H \rightarrow q_0\cdot Q^\ast$ given by
$h(w) = hide_{Q_A}(w)$. Let $G$ be the set of all strategies $g:
T_H \rightarrow Dist(\Sigma_I)$, and $F$ be the set of all strategies
$f: q_0\cdot Q^\ast\rightarrow Dist(\Sigma_I)$. Then $h$ can be lifted
to a bijection from $F$ to $G$ as follows: for $f\in F$, $g\in G$,
$h(f) = f\circ h$ and $h^{-1}(g) = g \circ h^{-1}$. 
Then $\mu_f(L) = \mu_{h(f)}(H\cap R)$  and $\mu_g(H\cap R) =
\mu_{h^{-1}(g)}(L)$. Thus $f\in F$ (resp. $g\in G$) is winning for the
environment iff $h(f)$  (resp. $h^{-1}(g)$) is winning for the
environment. \qed

Given a library $\L$ and monitor $A$, we can solve the embedded 
realizability problem for the augmented library $\L_A$ to obtain a regular
tree $T$, where $T = tree(C)$ for some composer $C$ over $\L_A$ such 
that $C$ satisfies $\alpha_A$. Then the tree 
$T' = xray(Q_A , wide_{Q_A}(tree(C)))$ is also regular, so 
$T' = tree(C')$ for some composer $C'$ over $\L$. Now we would like to 
use $C'$ to solve the DPW realizability problem, but $C'$ is only 
guaranteed to satisfy $A$ if $C$ is the augmentation of $C'$ by $A$. 
Therefore, to solve the DPW realizability problem, we have to obtain an 
automaton that accepts a tree $T'= tree(C')$ if the augmentation of $C'$ by 
$A$ satisfies $\alpha_A$.

\begin{theorem}\label{theorem:narrow}
Let $X$, $Y$ and $Z$ be finite sets. Given an  alternating automaton
$\mathcal{B}$ over $(Z\times Y)$-labeled $(X\times Y)$-trees, we can 
construct an alternating automaton $\mathcal{B'}$ over $Z$-labeled 
$X$-trees such that $\mathcal{B'}$ accepts a labeled tree 
$\tuple{X^\ast, V}$ iff $\mathcal{B}$ accepts 
$xray(Y, wide_Y(\tuple{X^\ast, V}))$. Further, $\mathcal{B}$ and 
$\mathcal{B'}$ have the same acceptance condition and 
$|\mathcal{B'}| = O(|\mathcal{B}|)$.
\end{theorem}

\proof
Let $\mathcal{B} = (Z\times Y, Q,\delta, q_0, \alpha)$ be an
alternating automaton that accepts $(Z\times Y)$-labeled $(X\times
Y)$-trees. We define automaton $\mathcal{B}_1 = (Z, Q\times Y,\delta',
(q_0, y_0),\alpha\times Y)$ over $Z$-labeled $(X\times Y)$-trees,
where for each $q\in Q$, $y\in Y$ and $z\in Z$, $\delta'((q,y),z)$ is
obtained from $\delta(q, (z,y))$ by replacing each atom $((x',y'),q')$
by the atom $((x',y'),(q', y'))$. So a state $(q,y)$ in
$\mathcal{B}_1$  corresponds to a state $q$ in $\mathcal{B}$ that
reads only nodes in direction $y$. Then $\mathcal{B}_1$ accepts a
$Z$-labeled $(X\times Y)$-tree $\tuple{(X\times Y)^\ast, V}$ iff
$\mathcal{B}$ accepts $xray(Y,\tuple{(X\times Y)^\ast, V})$. 

Next, we define alternating automaton $\mathcal{B'} = (Z, Q\times Y,
\delta'', (q_0, y_0), \alpha\times Y)$ over $Z$-labeled $X$-trees,
where for every $(q,y)\in Q\times Y$ and $z\in Z$, $\delta''((q,y),
z)$ is obtained from $\delta'((q,y), z)$ by replacing each atom
$((x,y'), (q', y'))$ by the atom $(x,q')$. Then  
for every $Z$-labeled $X$-tree $\tuple{X^\ast, V}$, we have
$\tuple{X^\ast, V}\in L(\mathcal{B'})$ iff $wide_Y(\tuple{X^\ast,
  V})\in L(\mathcal{B}_1)$ (See \cite{KV97c} for proof).  

Therefore, $\mathcal{B'}$ accepts $\tuple{X^\ast, V}$ iff
$\mathcal{B}$ accepts $xray(Y, wide_Y(\tuple{X^\ast, V}))$, and
$\mathcal{B'}$ is the required automaton. \qed

Given an alternating automaton $\mathcal{B}$, let $narrow_Y(\mathcal{B})$ 
denote the corresponding automaton constructed in Theorem 
\ref{theorem:narrow}.
\begin{theorem}
Let $\L$ be a library and $A$ be  
a monitor for $\L$. Then there exists an alternating parity tree 
automaton (APT) $\B$ such that, for all composers $C$ over $\L$, $\B$ accepts
$tree(C)$ iff $C$ satisfies $A$. Consequently, $\B$ is non-empty iff $\L$ 
realizes $A$.
\end{theorem}
\proof 
Let $A = (\Sigma_O, Q_A, s_0, \delta_A, \alpha_A)$.
Let $\mathcal{B'}$ be the NPT that accepts $tree(C')$ iff $C'$ satisfies
$\alpha_A$ and $C'$ is compatible with $R_A$, for all composers $C'$
over $\L_A$. Such a $\mathcal{B'}$ exists by Lemma \ref{lemma:embedded-DPW}. 
Let $\mathcal{B} = narrow_{Q_A}(\mathcal{B'})$. We show that
$\mathcal{B}$, which is an APT, is the required automaton. 

Let $C$ be a composer over $\L$. By Theorem~\ref{theorem:augmented},
$C$ satisfies $A$ iff $C_A$ satisfies $\alpha_A$. Therefore, $\B'$
accepts $tree(C_A)$ iff $C$ satisfies $A$. By Lemma
\ref{lemma:augmented},  \[tree(C_A) = xray(Q_A , wide_{Q_A}(tree(C)))\]
and by Theorem \ref{theorem:narrow}, 
$\B$ accepts a tree $T$ iff $\B'$ accepts $xray(Q_A ,
wide_{Q_A}(T))$. Thus, $\B$ accepts $tree(C)$ iff $C$ satisfies $A$.
Since an APT is nonempty iff it accepts a regular tree, and $\L$ realizes
 $A$ iff some composer $C$ over $\L$ satisfies $A$, therefore $\B$ is
non-empty iff $\L$  realizes $A$.
\qed

%\vspace{4pt}
Each transducer in the augmented library $\L_A$ has a set of final
states of size $|D||Q_A|$. Thus the automaton $\mathcal{B'}$ has size
exponential in both $|D|$ and $|Q_A|$. The translation from 
$\mathcal{B'}$ to $\mathcal{B}$ adds no blowup,
but $\mathcal{B}$ is an APT, while $\mathcal{B'}$ is an NPT.
Since emptiness for an alternating parity tree
automaton can be checked in time exponential in the size of the
automaton \cite{MS95}, therefore $\mathcal{B}$ can
be be checked for emptiness in time doubly exponential in $|D|$ and
$|Q_A|$.

%Since the DPW probabilistic realizability problem is a strict
%generalization of the control-flow synthesis problem that was shown to
%be 2EXPTIME-complete in \cite{LV09}, we have the following:

\begin{theorem}
The DPW probabilistic realizability problem is in 2EXPTIME. \qed
\end{theorem}

Again, if an alternating tree automaton is nonempty, then it must accept some
regular tree \cite{MS95}, and given a regular tree accepted by $\B$, we can
obtain a finite transducer that generates that tree. This transducer
is a composer that realizes $A$. Thus, we also obtain a solution to
the DPW probabilistic synthesis problem.

\begin{theorem}
The DPW probabilistic synthesis problem is in 2EXPTIME. \qed
\end{theorem}

%myv1: moved here.
The doubly exponential upper bound for our solution can be viewed
as follows: we inherit one exponential from the embedded parity solution
and the second exponential is introduced by the use of an APT to deal
with incomplete information. It is an open question whether the second
exponential can be avoided.

%\section{Probabilistic Composers}
%Here we consider a generalization of the problem where we allow the
%composer to also be a probabilistic transducer.

\section{Discussion and Future Work}

Component-based synthesis seeks to build systems that satisfy a given
specification using pre-existing components. This contrasts with
classical synthesis, where the aim is to build a system from
scratch. The component-based approach is closer in spirit to how
systems are built in the real world.
In this paper, we generalize the component-based synthesis problem to
a probabilistic setting. Our components are modeled as probabilistic
transducers and the specification is given as a deterministic parity
automaton. The composition
itself is described by a deterministic transducer, called a
\emph{composer}, which governs the transitions between components. 

We break the problem down in two
stages. First we solve a simpler version, which we call the
\emph{embedded parity synthesis problem}, where the specification is
embedded as parities in the components themselves. Our solution
combines techniques from Markov chain analysis and automata theoretic
verification. Then we show how to
solve the more general case of a separate specification, which we call
the \emph{DPW probabilistic synthesis problem}, by reducing
it to the simpler case using techniques from synthesis
with incomplete information. 

We show that the embedded parity synthesis problem is in EXPTIME and
the DPW probabilistic synthesis problem is in 2EXPTIME. The question
of tighter lower and upper bounds we leave for future work. In
particular, it is an open question whether the DPW probabilistic
synthesis problem is in EXPTIME. Another line of work is suggested by
the possibility of probabilistic composers. In recent work, we show that allowing the composer to be a probabilistic transducer makes the synthesis problem sensitive to the specification formalism \cite{NV12}. It turns out that probabilistic composers are more expressive than their deterministic counterparts for DPW specifications, but they have the same expressive power for embedded parity specifications.

%\bibliographystyle{plain}
%\bibliography{reactive}

%\end{document}

%\newpage
%\appendix
%
%\section{Proof of Theorem \ref{theorem:memoryless}}\label{appendix:memoryless}
%\renewcommand\thesection{\Alph{section}}
%
%

\end{document}